\documentclass{amsart}
\usepackage{graphicx}    
\usepackage{subcaption}  
\usepackage{wrapfig}		
\usepackage{amsmath,amssymb}
\usepackage{color}
\usepackage{url}

\usepackage{tikz}
\usepackage{tkz-euclide}

\title[Frank--Read dislocation source]{A quantitative model for the Frank--Read dislocation source based on pinned mean curvature flow}
\author{Thomas Hudson}
\author{Filip Rindler}
\author{Joshua Rydell}
\address{Mathematics Institute, University of Warwick, Coventry CV4 7AL, United Kingdom.}
\email{T.Hudson.1@warwick.ac.uk}
\email{F.Rindler@warwick.ac.uk}
\email{Josh.Rydell@warwick.ac.uk}
\date{August 2024}
\dedicatory{In memory of G\"{u}nter von H\"{a}fen}

\def\bfb{\boldsymbol{b}}
\def\bfe{\boldsymbol{e}}
\def\bff{\boldsymbol{f}}
\def\bfn{\boldsymbol{n}}
\def\bft{\boldsymbol{t}}
\def\bfsig{\boldsymbol{\sigma}}
\def\bfphi{\boldsymbol{\varphi}}
\def\bfpsi{\boldsymbol{\psi}}
\def\bfv{\boldsymbol{v}}

\usepackage{algpseudocode}
\usepackage{algorithm}

\def\dd{\mathrm{d}}
\def\ds{\dd s}
\def\dt{\dd t}

\newcommand{\brac}[1]{\left(#1 \right)}
\newcommand{\set}[1]{\left\{ #1 \right\}}

\def\bbR{\mathbb{R}}
\newcommand{\ip}[2]{\left \langle #1 , #2 \right \rangle}

\begin{document}

\begin{abstract}
    This work introduces a simple quantitative model for the Frank--Read source, considered to be one of the most important micro-mechanical mechanisms of dislocation creation in crystalline materials. It has long been known that these sources create dislocations in a repetitive, oscillatory process, which is driven by an external shear force. Unlike the existing explanations in the literature, the model introduced in the present article is based on just a few simple physical principles, namely line tension and dislocation motion due to a single slip plane flow rule, together with a pinning constraint on the ends of the central dislocation line. A complete discretisation, including suitable re-meshing and ``topological cutting'' algorithms, is described and simulation results are discussed. Despite its conceptual simplicity, the model and discretisation described in the present work yield remarkably accurate predictions about the shape and properties of the Frank--Read source. In particular, it is shown that only one dimensionless parameter controls the dynamics of the Frank--Read source if one neglects crystal anisotropy. This allows to derive an emergent law about the length of dislocation line generated per shear energy.
\end{abstract}

\maketitle

\section{Introduction}
\label{sec:intro}

It is well-established that stress applied to metals leads to dislocation multiplication and flow, thereby enabling their ductility~\cite{AndersonHirthLothe17book,HullBacon11book}. The Frank--Read source \cite{FR50} is one prominent phenomenon by which new dislocation loops are believed to be generated in the early onset of plastic deformation. At such sources, stress drives dislocations to repeatedly self-intersect, causing topological change and dislocation multiplication. Such behaviour has been observed experimentally in many different materials; see for example \cite{Marshall1975,Geipel1996}. However, the evolution in time of the Frank--Read mechanism is difficult to observe due to the challenges of both imaging dislocations under strain and tracking their motion directly. Thus, one must resort to mathematical and numerical models to provide quantitative predictions of the behaviour of dislocation multiplication at the microscopic scale.

\begin{figure}[tp]
\begin{subfigure}{.45\textwidth}
  \centering
  \includegraphics[height=30mm]{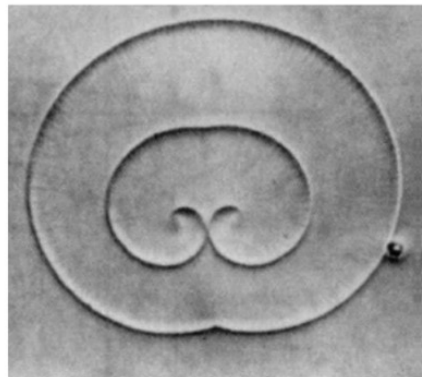}
  \caption{Experimental image.} 
\end{subfigure}\hfill
\begin{subfigure}{.45\textwidth}
  \centering
  \includegraphics[height=36mm]{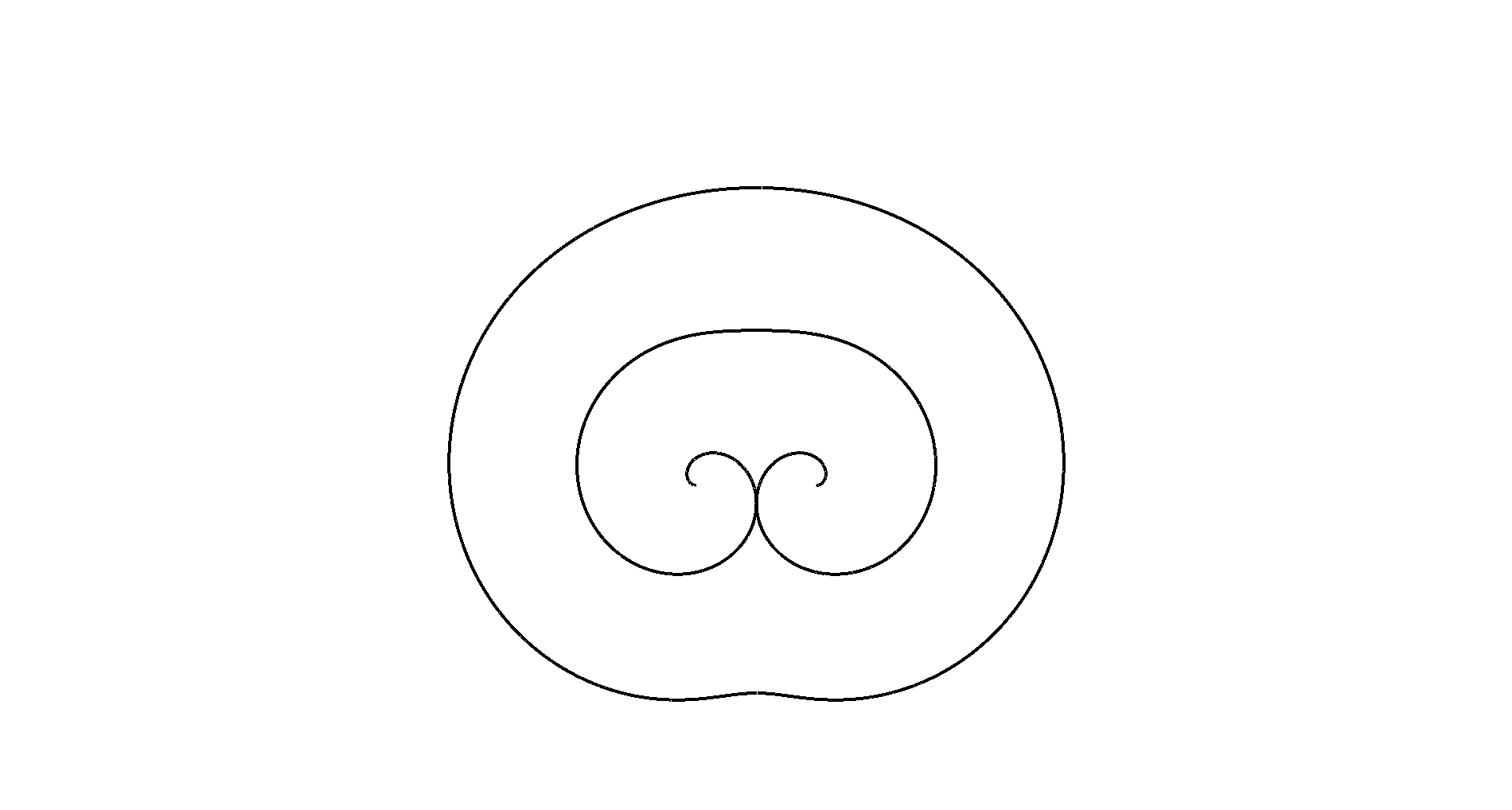}
  \caption{Simulated source.} 
\end{subfigure}
    \caption{Comparison between experimental image of a Frank--Read source in NiFe, adapted from Figure~1.2.28 of \cite{Tad24}, and a simulated curve using our model. In (b), the dimensionless parameter is $f_s = 5 \times 10^{-3}$, and $t=0.61$.}
    \label{fig:experiment}
\end{figure}

Our main contributions in this work is to provide a self-contained treatment of the Frank-Read source under minimal physical assumptions. In particular:
\begin{itemize}
    \item We derive a simple model of dislocation motion at a Frank--Read source under the assumption that core energies are independent of dislocation character, and these contributions dominate long-range elastic interactions;
    \item We non-dimensionalise the resulting model, allowing us to reduce the model to just one free dimensionless parameter; and
    \item We then simulate the model, describing an open source numerical implementation provided, and demonstrate that our approach indeed reproduces experimental results: see Figure~\ref{fig:experiment}.
\end{itemize}
It is a remarkable consequence of our modelling approach that a complex phenomenon like the Frank--Read source can be accurately described by a very simple set of assumptions, which elegantly reduce the prediction thereof to a geometric flow problem.

Further, in contrast to existing Discrete Dislocation Dynamics (DDD) approaches \cite{BC06}, the force computation required to update dislocation positions becomes local in space, and so much less computationally costly to simulate. We emphasise that we do not expect the assumptions we make to be valid in all possible materials and modelling situations, but our results nevertheless serve as a benchmark to be compared with more sophisticated approaches. We hope that clarifying the assumptions of the mathematical theory we outline will enable further investigation of the basic assumptions underlying DDD, thereby improving the robustness of computational modelling approaches.

\begin{figure}[tbp]
    \centering
    \includegraphics[width=1\linewidth]{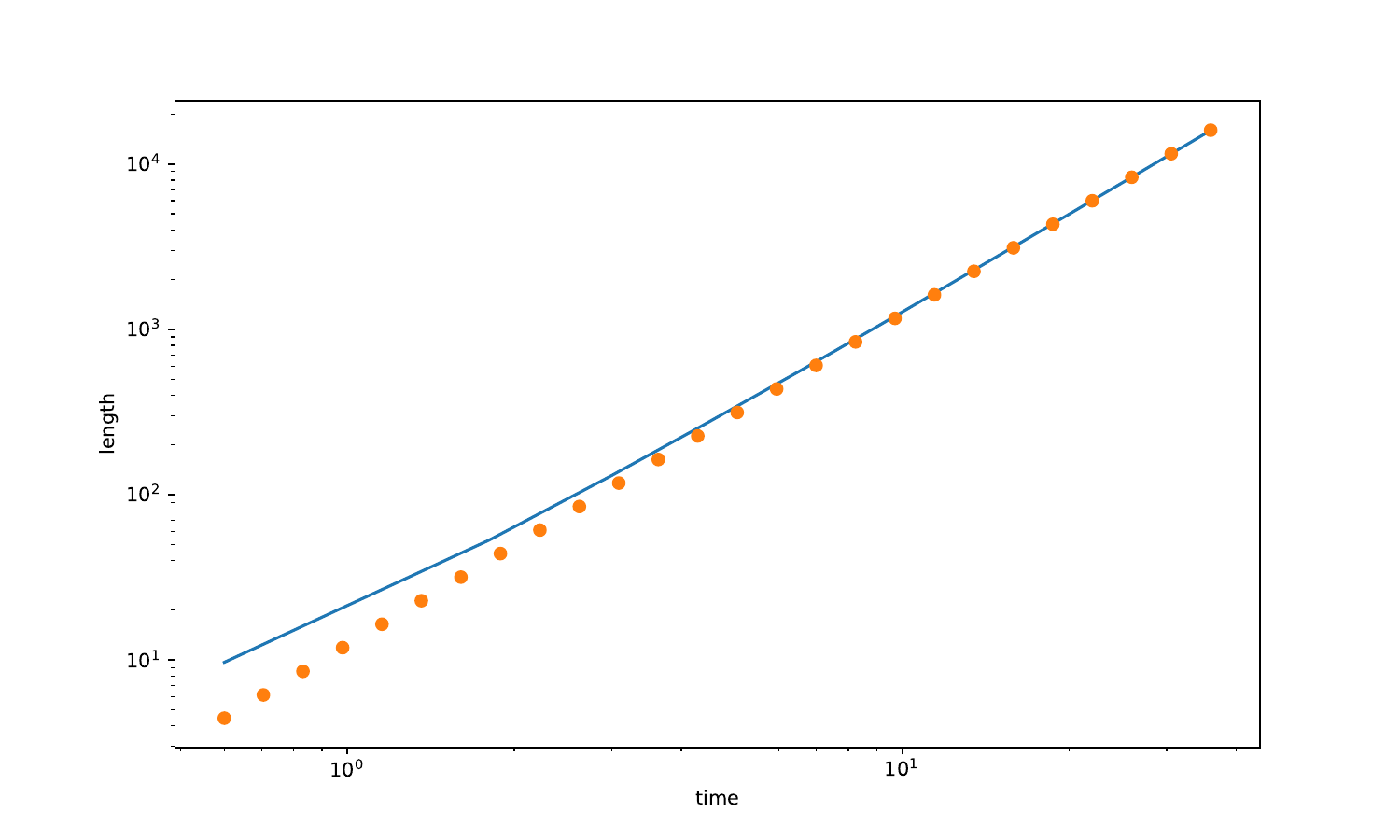}
    \caption{Log-log plot of the total length of the dislocation lines generated by a constant applied shear stress in a simulation (solid line); a fitted asymptotic quadratic rate is shown for reference (dotted line). The simulation used dimensionless parameter $f_s = 5 \times 10^{-3}$, and time step $\tau = 1.2 \times 10^{-3}$.}
    \label{fig:length}
\end{figure}

Our model enables us to derive precise laws for the Frank--Read source, most notably the following \emph{square law} of the length of generated dislocations under a constant applied stress: The total length $\ell$ of generated dislocations is proportional to the square of the elapsed time $t$, namely
\[
  \ell \propto t^2,
\]
at least to leading order. This law can be directly observed in Figure~\ref{fig:length}, and an analytic justification of this law based on our model can be found in Section~\ref{sec:results}.

The remainder of the paper is structured as follows. In Section~\ref{sec:derivation} we give some background on the mathematical model, namely pinned and forced mean curvature flow, before presenting a derivation of our model for dislocation motion. Then, Section~\ref{sec:discretisation} describes our approach to numerically discretising the model, and Section~\ref{sec:results} presents and analyses the results of our simulations. Finally, Section~\ref{sec:conclusion} summarises our conclusions.

\subsection*{Acknowledgments}

We would to thank Paolo Bonicatto and Giacomo Del~Nin for discussions related to this work. This project has received funding from the European Research Council (ERC) under the European Union's Horizon 2020 research and innovation programme, grant agreement No 757254 (SINGULARITY). Following the convention in the theoretical sciences, the authors are ordered alphabetically.

\section{Model}
\label{sec:derivation}

This section derives an analytic model for the classic Frank--Read source for dislocation multiplication. We consider a single active slip plane in an elastic crystal of infinite extent, thereby mimicking the behaviour in a bulk crystal and avoiding the need to consider interactions with free surfaces or grain boundaries. We then model the dislocations as being subject to a configurational force balance, in which dislocation evolution is driven by the action of an externally applied stress. We will see that our approach yields a form of pinned and forced mean curvature flow.

\smallskip\noindent\textbf{Classical mean curvature flow.}
Mean curvature flow (MCF), also called curve shortening flow when only considering curves and not higher-dimensional surfaces, is a form of geometric flow in which curves move in the direction of their curvature vector (which has magnitude equal to the inverse of the radius of curvature radius and points towards the centre of the osculating circle). In subsequent sections, we will show that we may model dislocation motion as a variant of MCF where a forcing term and pinning is included. For the reader's convenience, we here provide some background on known results concerning the properties of MCF.

Consider a regular embedded oriented curve $\gamma$ in $\bbR^2$. If sufficiently smooth, the curve has a signed curvature, unit tangent vector, and unit normal vector, which are denoted $k$, $\bft$ and $\bfn$ respectively. As usual for a plane curve, $\bfn$ can be obtained by rotating the unit tangent vector $\bft$ anticlockwise by $90^\circ$, and the product $k \bfn$ will be referred to as the curvature vector. Note that the sign of $k$ is such that $k \bfn$ points towards the centre of the curvature circle attached to the curve at a given point.

A single time-dependent curve $\gamma(t)$ moves by MCF if the velocity $\bfv$ of each point on the curve at each time $t$ is equal to the curvature vector, i.e. at all points on the curve (except possibly its ends) we require that
\begin{equation}\label{MCF}
    \bfv=k\bfn.
\end{equation}
More generally, a time-dependent family of curves moves by MCF if every connected component moves by MCF in the sense described.

It turns out that MCF can be viewed as a gradient flow of the length functional of the curve (see \cite{Cm11}); in particular, we have 
\begin{equation}\label{eq:length_derivative}
\frac{\dd}{\dt}\text{Length}\big(\gamma(t)\big) = - \int_{\gamma(t)} \ip{\bfv}{k\bfn} \,\ds.
\end{equation}
From here, it is possible to observe that the length of the curve is decreasing under this flow, since
\[
\frac{\dd}{\dt} \text{Length}(\gamma(t)) = \frac{\dd}{\dt}\int_{\gamma(t)}\,\ds =-\int_{\gamma(t)} k\ip{k\bfn}{\bfn}\dd s = -\int_{\gamma(t)} k^2 \dd s\leq 0.
\]
Moreover, curves evolving under MCF do so independently of any particular parametrisation, and isometries do not affect the curve's shape under this evolution. Short-time existence for MCF was proved in \cite{GH86} for the evolution beginning from any smooth, closed, embedded curve.

\smallskip\noindent\textbf{Parametric form of MCF.}
We will only consider closed curves from now on. If such a family of curves is parametrised by
\[
  \bfphi = \bfphi(\theta,t) :  [0,2\pi] \times [0,T) \to \bbR^2,
\]
where the first argument corresponds to arc length and the second one to time, then it is possible to express the curvature vector as
\[
k\bfn = \frac{1}{|\bfphi^0_\theta|} \left (\frac{\bfphi^0_\theta}{|\bfphi^0_\theta|} \right )_\theta.
\]
Here, the index ``$_\theta$'' denotes the $\theta$-derivative. Similarly, the velocity of a point on the curve $\bfv$ may be expressed as the time derivative of the parametrisation, denoted $\bfphi_t$. Combining these representations, \eqref{MCF} is equivalent to requiring that the parametrisation $\bfphi$ satisfies the following partial differential equation:
\begin{equation} \label{MCF_parametric}
\bfphi_t = \frac{1}{|\bfphi_\theta|} \left (\frac{\bfphi_\theta}{|\bfphi_\theta|} \right )_\theta.
\end{equation}
In order to specify the evolution properly, this equation must be supplemented with an initial condition and boundary conditions. In the case of a closed curve, we require that
\begin{equation}\label{eq:MCF_periodicBCs}
\begin{cases}
\bfphi(\theta,0) = \bfphi^0(\theta) & \text{for all $\theta$ in } [0,2\pi], \\
\bfphi(\theta + 2\pi , t) = \bfphi(\theta, t) & \text{for all } (\theta, t) \in  \bbR \times (0,T),
\end{cases}    
\end{equation}
and together, \eqref{MCF_parametric} and \eqref{eq:MCF_periodicBCs} specify a parametric form of MCF.

\smallskip\noindent\textbf{Geometric setup.}
Since we consider dislocations undergoing straightforward glide motion without cross-slip, dislocation curves are confined to a single slip plane. We therefore need only consider the motion in two dimensions from now on, ignoring the third spatial dimension.

At initial time, we assume a single segment of dislocation line connects two pinning sites. By scaling coordinates appropriately, we are free to require that these sites are at $(x,y) = (\pm 1,0)$; see Figure~\ref{fig:setup}. We assume that this segment of dislocation line has Burgers vector $\bfb$ which, due to physical constraints on glide motion, must also lie in the same slip plane (see for example the discussion in Chapter~1 of \cite{HL82}).

\tikzset{
  on each segment/.style={
    decorate,
    decoration={
      show path construction,
      moveto code={},
      lineto code={
        \path [#1]
        (\tikzinputsegmentfirst) -- (\tikzinputsegmentlast);
      },
      curveto code={
        \path [#1] (\tikzinputsegmentfirst)
        .. controls
        (\tikzinputsegmentsupporta) and (\tikzinputsegmentsupportb)
        ..
        (\tikzinputsegmentlast);
      },
      closepath code={
        \path [#1]
        (\tikzinputsegmentfirst) -- (\tikzinputsegmentlast);
      },
    },
  },
  mid arrow/.style={postaction={decorate,decoration={
        markings,
        mark=at position .5 with {\arrow[#1]{stealth}}
      }}},
}

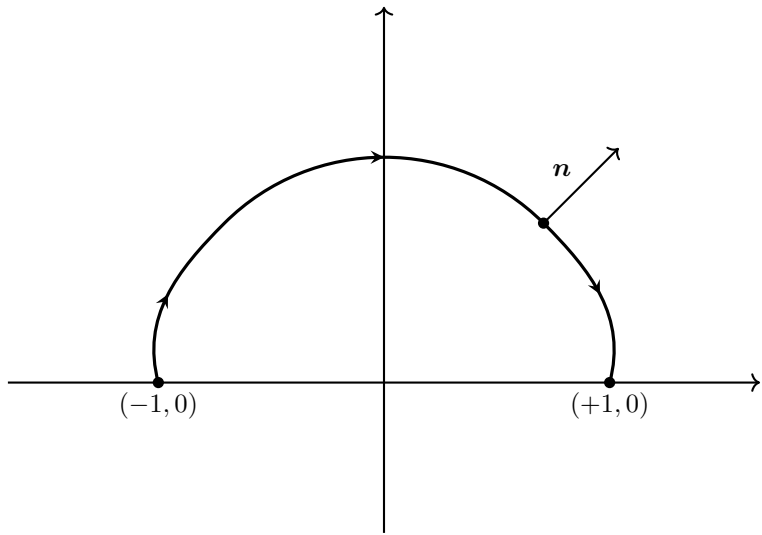
\begin{figure}[tbp]\centering
\begin{tikzpicture}
\usetikzlibrary{decorations.markings}

\draw[thick,->] (-5,0) -- (5,0);
\draw[thick,->] (0,-2) -- (0,5);
\tkzDefPoint(-3,0){A};
\tkzDefPoint(+3,0){B};
\tkzDefPoint(-2.121,2.121){C};
\tkzDefPoint(+2.121,2.121){D};
\tkzLabelPoint[below](A){$(-1,0)$};
\tkzLabelPoint[below](B){$(+1,0)$};
\node at (A)[circle,fill,inner sep=1.5pt]{};
\node at (B)[circle,fill,inner sep=1.5pt]{};
\node at (D)[circle,fill,inner sep=1.5pt]{};
\path[draw,postaction={on each segment={mid arrow=black}},very thick] (A) to[out=105,in=-135] (C) to[out=45,in=135] (D) to[out=-45,in=75] (B);
\draw[thick,->] (D) -> (3.121,3.121) node[midway,above left] {$\bfn$};
\end{tikzpicture}
\caption{Illustration of the geometric setup.}
\label{fig:setup}
\end{figure}

\smallskip\noindent\textbf{Internal energy.}
We model dislocations as being subject to line tension forces, which are configurational forces induced by the energetic cost which must be expended to form dislocation cores at the atomistic scale. As mentioned in the introduction, a novelty in our approach is that we choose to neglect long-range elastic interactions, focusing instead on the case where there is a dominant balance between line tension forces and the externally applied forces. While this yields a less complex model than the approach usually take in dislocation dynamics simulations \cite{BC06}, Figure~\ref{fig:experiment} indicates that our model is nevertheless compatible with experimental observations. This suggests that the assumption of weak elastic interaction is valid in some material systems. We refer to~\cite{ContiGarroniOrtiz15,Ginster19} and the references cited therein for rigorous justifications of the line tension energies as multiples of the (potentially anisotropic) dislocation line length.

To derive a force balance, we begin by considering the energy of the material system. We will suppose that the internal energy of the body in the region of interest arises mainly due to the presence of dislocations, and as such the internal energy is expressed solely in terms of the dislocations present. Implicitly, this assumption entails that the release of any stored elastic energy happens on a much faster timescale than the plastic evolution we are considering: this is a common assumption in many dislocation dynamics models \cite{HL82,BC06}.

Suppose therefore that $E_0$ is the formation energy of dislocation line per unit length, sometimes also termed the `core energy' of the dislocation; then the total line tension energy of the dislocation is
\[
\mathcal{E}(\gamma) = \int_\gamma E_0\,\ds.
\]
In practice, the core energy of a dislocation is challenging to determine, as any measurement of it inherently requires an artificial separation of core and elastic energy. The simplest possible modelling assumption we can make is that the core energy is completely independent of the local character, i.e. the relative orientation of the line and the Burgers vector, which determines whether the dislocation is screw, edge or of mixed type. The approximate constancy of the core energy is a phenomenon that is observed in practice in some materials: in \cite{BCAB21}, it is calculated that for a fixed Burgers vector, the core energies in various BCC metals vary with orientation by around 10--20\%.

Under this assumption $E_0$ is constant, and in this case, we have
\[
\mathcal{E}(\gamma) = E_0 \cdot \,\text{Length}(\gamma).
\]
It is this particular case that we will choose to focus on for the remainder of this work.

The configurational force acting on the dislocation due to line tension can be seen as the negative gradient of this energy, $-\nabla \mathcal{E}(\gamma)$, expressible as
\[
\langle-\nabla \mathcal{E}(\gamma),\bfpsi\rangle = \int_\gamma E_0\,k\bfn\cdot \bfpsi\,\ds.
\]
Here $\bfpsi$ is an appropriate variation, and we recall from above that $k$ is the curvature and $\bfn$ is the unit normal vector. As such, we see that the line tension forces are expressible as
\[
\bff^{\text{int}} = E_0\, k\bfn.
\]

\smallskip\noindent\textbf{Dissipative forces.}
We assume that dissipative forces act to oppose the motion of dislocations, and that these may be treated as an isotropic viscous friction on the slip plane. This entails that the dissipative force per unit length acting on this dislocation line is
\[
\bff^{\text{diss}} = -\mu\bfv.
\]
Here, as above, $\bfv$ is the velocity of the curve, and $\mu$ is a coefficient of viscous friction which has units of dissipated power per unit area. For context, a similar assumption is often made in the literature on dislocation dynamics, where it is often assumed that the coefficient $\mu$ can be written as
\[
\mu = \frac{b}{B}
\]
where $b$ is the length of the Burgers vector of the dislocation and $B$ is a mobility coefficient; see \cite{BC06} for further discussion.

We remark that the resulting dynamics are \emph{rate-dependent}. If one were to use rate-independent dynamics, as in~\cite{HudsonRindler22}, the above relation between dissipative force and velocity would instead have to be modelled using a differential inclusion, often also called a Biot inclusion. This is a relation of the form
\[
-\bff^{\text{diss}} \in \partial R(\bfv),
\]
where $\partial R$ is the subdifferential of a convex \emph{dissipation potential} $R$, assumed to be positively $1$-homogeneous (so that $R(\lambda \bfv) = \lambda R(\bfv)$ for all $\lambda \geq 0$). Written out, this relation means that
\[
  R(\bfv) - \bff^{\text{diss}} \cdot (\bfv' - \bfv) \leq R(\bfv')
\]
for all velocities $\bfv'$. If instead one took $R(\bfv) = -2\mu |\bfv|^2$, we would recover the above relation $\bff^{\text{diss}} = -\mu\bfv$. Which kind of dynamics to choose is clearly a modelling choice (see, e.g.,~\cite{Fremond02,Mielke03a}). In this work we are mostly interested in the shape of the dislocations in the Frank--Read source, and less on the precise dynamics in time, so we stick with the rate-dependent model, which is much easier to implement numerically.

\smallskip\noindent\textbf{External forcing.}
In early thereotical work on dislocations \cite{PK50}, an energy argument was used to establish that dislocations experience configurational forces in the presence of a stress field in the material. In particular, it was shown that the configurational force per unit length on a dislocation acting due to applied stress is expressed as
\[
\bff^{\text{ext}} = (\bfsig\,\bfb)\times\bft,
\]
where $\bfsig$ is the external stress field; $\bfb$ is the Burgers vector; and as above, $\bft$ is the unit tangent vector which orients the dislocation line.

In the context where we consider a single active slip plane, we can be more precise.
If a uniform stress state $\bfsig$ is induced within the material with Cartesian components $\sigma_{ij}$, the Burgers vector is $\bfb= b\bfe_1$ and the tangent $\bft = \cos\theta\,\bfe_1+\sin\theta\,\bfe_2$, a short calculation shows that
\[
(\bfsig\bfb)\times\bft = -b\,\sigma_{13}\sin\theta\,\bfe_1+b\,\sigma_{13}\cos\theta\,\bfe_2+b(\sigma_{11}\sin\theta-\sigma_{12}\cos\theta)\bfe_3.
\]
In practice, only the projection of this force onto the slip plane matters for the motion of a dislocation undergoing glide, since motion out of plane is prevented by incompressibility, which we henceforth assume. With the geometric setup considered here, this projection entails that the $\bfe_3$ component of the configurational force is irrelevant. This means that we can write the force per unit length projected onto the slip plane as
\[
\bff^\text{ext} = b\sigma_{13}(-\sin\theta\,\bfe_1+\cos\theta\bfe_2) = b\sigma_{13}\bfn
\]
where $\bfn$ is the normal vector to the dislocation line. As such, we see that the driving force induced by a general stress state induces a force which depends only on the shear stress induced on the slip plane.

Parenthetically, we observe that the choice to choose $\bfb = b\bfe_1$ is not truly restrictive: if the Burgers vector was instead $\bfb = b(\cos\phi,\sin\phi)$ for some polar angle $\phi$, we would have
\[
\bff^\text{ext} = b \big(\sigma_{13} \cos\phi + \sigma_{23}\sin\phi\big)\bfn.
\]
In this case the bracketed terms in the expression above would play the role of $\sigma_{13}$ in the calculation below.

\smallskip\noindent\textbf{Force balance.}
Assuming that dislocations undergo quasistatic evolution, so that the configurational forces acting on the dislocation line are balanced at all times, we must have that
\[
\bff^{\text{diss}}+\bff^{\text{int}}+\bff^{\text{ext}}=\boldsymbol{0}.
\]
Substituting the expressions derived above and rearranging, we arrive at the dimensional equation
\begin{equation*}
\bfv=\left(\frac{b\sigma_{13}}{\mu}+\frac{E_0}{\mu}k\right)\bfn.
\end{equation*}
Nondimensionalising this equation, we may choose units such that
\begin{equation}\label{eq:FR}
\tilde{\bfv}=\big(f_s+\tilde{k}\big)\bfn,
\end{equation}
where $\bfv = V\tilde{\bfv}$ and $k = \frac{1}{L}\tilde{k}$, with a characteristic velocity scale $V$ and characteristic radius of curvature $L$ defined to be
\[
V = \frac{b\sigma_{13}}{\mu f_s} \qquad \text{and}\qquad L = \frac{E_0 f_s}{ b\sigma_{13}}.
\]
We will refer to the remaining dimensionless free parameter $f_s$ as the `forcing scalar'; this parameter can be seen as describing the relative magnitude of the external forcing versus the line tension forces. For notational simplicity, we will drop tildes from the dimensionless variables from now on.

In parametric form, the dimensionless equation \eqref{eq:FR} can be expressed as
\begin{equation}\label{eq:FR_parametric}
\bfphi_t = f_s \frac{\bfphi_\theta^\perp}{|\bfphi_\theta|}+\frac{1}{|\bfphi_\theta|} \left (\frac{\bfphi_\theta}{|\bfphi_\theta|} \right )_\theta,
\end{equation}
where $\bfphi_\theta^\perp$ denotes the vector obtained by rotating $\bfphi_\theta$ anticlockwise by $90^\circ$.

To complete our description of the evolution problem for the Frank--Read source, we prescribe an initial condition and Dirichlet boundary conditions pinning the dislocation line at $(\pm 1,0)$, i.e.
\begin{equation}
\label{eq:FR_BCs}
\begin{cases}
\bfphi(\theta,0) = \bfphi^0(\theta) & \text{for all $\theta$ in } [0,2\pi], \\
\bfphi(0 , t) = (-1,0) & \text{for all } t \in   (0,T), \\
\bfphi(2 \pi , t) = (+1,0) & \text{for all } t \in   (0,T).
\end{cases}
\end{equation}
In our simulations, we will choose the initial condition to be a straight line segment between the pinning points.

\smallskip\noindent\textbf{Remarks.}
Equation \eqref{eq:FR} can be seen as a variant of MCF for the dislocation lines, where there is an additional driving force of fixed magnitude that always points in the normal direction. In the limit where $f_s\to 0$, it reduces to exactly this case.
It is immediate to see from \eqref{eq:FR} that there is a competition between the effect of the external stress, captured by the forcing term, and the relaxation of the internal energy, captured by the curvature term. In particular, any steady states which exist are given by curves of constant curvature with
$k = -f_s$: these must be circular arcs with a radius which is the inverse of this quantity, i.e.
$r_* =-f_s^{-1}$.

\section{Numerical discretisation}
\label{sec:discretisation}
In this section, we describe the methodology used to approximate solutions to the model of dislocation motion \eqref{eq:FR}. We adopt a finite difference approach, although the resulting method is closely connected to the finite element approach discussed in \cite{Dec05}, and the regularised method of lines approach discussed in \cite{PB09,PBK10}.

\smallskip\noindent\textbf{Numerical approximation of MCF.}
As with many PDE problems, analytic solutions to the mean curvature flow are limited to very simple cases, and hence we must resort to numerical approximations. A review of the existing finite-element based approaches to simulating MCF for smooth closed curves is presented in Section~4.1 of \cite{Dec05}. Here, we briefly summarise this discussion for the reader's convenience. In \cite{D94}, it was shown that (under appropriate conditions) such solutions $\bfphi$ of MCF can be approximated by finite-element approximations $\bfphi_h$ which satisfy the error estimates
\[\max_{[0,T]} \| \bfphi - \bfphi_h \|_{L^2([0,2\pi])} + \left (\int^T_0 \|\bfphi_\theta - \bfphi_{h\theta}  \|^2_{L^2([0,2\pi])}\,\dt \right )^{1/2} \leq ch , \]
\[\max_{[0,T]} \| \bfphi_t - \bfphi_{th} \|_{L^2([0,2\pi])} + \left ( \int^T_0 \|\bfphi_{t\theta} - \bfphi_{ht\theta}  \|^2_{L^2([0,2\pi])}\,\dt \right )^{1/2} \leq ch , \]
where the constant $c$ depends only on $\bfphi$ and $T$.

These results provide convergence guarantees and rates for a finite element method in the context of MCF starting from an initial condition which is a closed, embedded curve. As we will see, the model for the dislocation motion we derive in the following section bears similarities to the MCF problem described here, albeit with an additional forcing term which depends upon the normal vector to the curve at each point. For our model, adopting a finite-element approach turns out to lead to terms which depend upon the parametrisation. Instead, we propose a geometrically intrinsic finite difference approach. While we expect similar error analysis to hold for the scheme we propose, this lies beyond the scope of the present work.

\smallskip\noindent\textbf{Spatial discretisation.} 
Defining the internodal length $q_j:=|\bfphi_{j}-\bfphi_{j-1}|$ for each $j$, we approximate the terms on the right-hand side of \eqref{eq:FR_parametric} as follows:
\begin{align*}
\frac{1}{|\bfphi_\theta|}\left(\frac{\bfphi_\theta}{|\bfphi_\theta|}\right)_\theta &\approx \frac{1}{\frac12(q_{j+1}+q_{j})}\left(\frac{\bfphi_{j+1}-\bfphi_j}{q_{j+1}}-\frac{\bfphi_{j}-\bfphi_{j-1}}{q_{j}}\right)\\
\text{and}\quad f_s\frac{\bfphi_\theta^\perp}{|\bfphi^\perp_\theta|}&\approx f_s \frac{\tfrac12(\bfphi_j-\bfphi_{j-1})^\perp+\tfrac12(\bfphi_{j+1}-\bfphi_{j})^\perp}{\big|\tfrac12(\bfphi_j-\bfphi_{j-1})^\perp+\tfrac12(\bfphi_{j+1}-\bfphi_{j})^\perp\big|}.
\end{align*}
We note that an advantage of this approach is that both of these approximations can be thought of as geometrically intrinsic: they do not depend explicitly on the choice of the underlying parametrisation; this would not be the case for a finite element approach.

Nevertheless, if we do introduce the discretisation step length $h$ in the domain of a paremetrisation $\bfphi:[0,2\pi]\to\bbR^2$, we can see that
\begin{align*}
    \tfrac12(q_j+q_{j+1}) &= |\bfphi_\theta(\theta_j)|h+O(h^2),\\
    \frac{\bfphi_{j+1}-\bfphi_j}{q_{j+1}}&=\frac{\bfphi'(\theta_j)h+\frac12\bfphi''(\theta_j)h^2+O(h^3)}{|\bfphi'(\theta_j)h+\frac12\bfphi''(\theta_j)h^2+O(h^3)|}\\
    -\frac{\bfphi_{j}-\bfphi_{j-1}}{q_{j}}&=\frac{\bfphi'(\theta_j)h-\frac12\bfphi''(\theta_j)h^2+O(h^3)}{|\bfphi'(\theta_j)h-\frac12\bfphi''(\theta_j)h^2+O(h^3)|}.
\end{align*}
By further carefully Taylor expanding, we can establish that
\[
\frac{1}{\tfrac12(q_j+q_{j+1})}\left(\frac{\bfphi_{j+1}-\bfphi_j}{q_{j+1}}-\frac{\bfphi_{j}-\bfphi_{j-1}}{q_{j}}\right) = \frac{\bfphi''(\theta_j)}{|\bfphi'(\theta_j)|^2}-\frac{\bfphi''(\theta_j)\cdot\bfphi'(\theta_j)}{|\bfphi'(\theta_j)|^3}+O(h^2),
\]
where we note that terms from the expansion of the terms in parentheses containing even powers of $h$ cancel exactly.
For the forcing term, we note that the approach taken involves a centred finite difference, yielding
\[
\tfrac12(\bfphi_{j+1}-\bfphi_{j-1})^\perp = \bfphi'(\theta_j)^\perp h +O(h^3),
\]
and hence 
\[
 f_s \frac{\tfrac12(\bfphi_j-\bfphi_{j-1})^\perp+\tfrac12(\bfphi_{j+1}-\bfphi_{j})^\perp}{\big|\tfrac12(\bfphi_j-\bfphi_{j-1})^\perp+\tfrac12(\bfphi_{j+1}-\bfphi_{j})^\perp\big|}
 \approx f_s\frac{\bfphi_\theta^\perp}{|\bfphi^\perp_\theta|}+O(h^2).
\]
From these results, we observe that this means that the spatial discretisation chosen is (locally) second-order accurate.

\smallskip\noindent\textbf{Time discretisation.}
We next discretise in time. For simplicity, and due to the need to remesh regularly, discussed further below, we choose to adopt a single step scheme, inspired by that proposed in Section~4.1 of \cite{Dec05} for MCF. We fix a time step $\tau$ and define discrete times $t_m = m\tau\in[0,T]$, approximating time derivatives by forward differences
\[\bfphi_{ht}(\cdot,t_m) \approx \frac{\bfphi^{m+1}_h - \bfphi^m_h}{\tau}\]
where $\bfphi^m_h$ approximates $\bfphi_h(\cdot,t_m)$. For numerical stability, we choose to treat the forcing terms implicitly in time, so that we must iteratively solve
\begin{multline}
     \label{lumped_fully_discrete_FE_MCF}
\frac 1 {\tau} (\bfphi^{m+1}_j - \bfphi^m_j)  =\frac{2}{ q^m_j+q^m_{j+1}}\left(\frac{\bfphi^{m+1}_{j+1}-\bfphi^{m+1}_j}{q^m_{j+1}} - \frac{\bfphi^{m+1}_{j}-\bfphi^{m+1}_{j-1}}{q^m_j}\right)\\
+
f_s \frac{(\bfphi^{m+1}_{j+1}-\bfphi^{m+1}_{j-1})^\perp}{|\bfphi^{m+1}_{j+1}-\bfphi^{m+1}_{j-1}|}
\end{multline}
for all indices $m$.
In practice, (\ref{lumped_fully_discrete_FE_MCF}) may be recast as a pair of block tridiagonal linear systems for the $x$ and $y$ coordinates of the nodes. These can be solved efficiently using the Thomas algorithm; for further details, see for example \cite{Bat06}.

\smallskip\noindent\textbf{Boundary conditions.}
In the case of the Frank--Read source, we assume dislocations lines are pinned at their endpoints, and so Dirichlet boundary conditions must be introduced to describe this. The resulting problem becomes:
\begin{equation} \label{pinned_MCF}
\begin{cases}
\bfphi_t - \frac{1}{|\bfphi_\theta|} \left (\frac{\bfphi_\theta}{|\bfphi_\theta|} \right )_\theta = 0,  & \text{in } (0,2\pi) \times (0,T), \\
\bfphi(\theta,0) = \bfphi^0(\theta) & \text{for $\theta$ in } [0,2\pi], \\
\bfphi(0 , t) = \bfphi^0(0) & \text{for all } t \in   (0,T), \\
\bfphi(2 \pi , t) = \bfphi^0(2 \pi) & \text{for all } t \in   (0,T), \\
\end{cases}
\end{equation} 
Discretising this system gives rise to a discrete system with matrices
\begin{equation*}
\begin{gathered}
 A_\text{pinned}^m = 
\begin{pmatrix}
-\frac 1 {q_2^m} &c^m_{2} & -\frac 1 {q_3^m} &  & 		 \\
 & \ddots & \ddots & \ddots & & 				\\
 & & -\frac{1}{q^m_{N-1}} & c^m_{N-1}&-\frac{1}{q^m_{N}} 
\end{pmatrix}
,\\\bfphi^m_{\bullet, \text{ pinned}} = 
\begin{pmatrix}
\bfphi^m_{2,\bullet} \\[3mm]
\vdots \\[3mm]
\bfphi^m_{N-1, \bullet} \\
\end{pmatrix}\quad\text{and}\quad
B^m_{\bullet, \text{ pinned}} = \begin{pmatrix}
b_{2, \bullet}^m  + \bfphi^m_{1, \bullet} / q^m_1 \\
b_{3, \bullet}^m \\
\vdots \\
b^m_{N-2, \bullet}\\
b_{N-1, \bullet}^m +\bfphi^m_{N, \bullet} / q^m_{N} 
\end{pmatrix}.
\end{gathered}
\end{equation*}

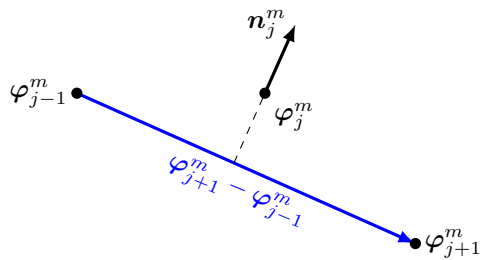
\begin{figure}[tbp]
    \centering
    \begin{tikzpicture}
    \tikzset{>=latex}
        \coordinate (A) at (2,-2);
        \coordinate (B) at (0,0);
        \coordinate (Bmean) at (-0.412, -0.928);
        \coordinate (Bnorm) at (0.406,0.914);
        \coordinate (C) at (-2.5,0);

        \node[right] at (A) {$\bfphi_{j+1}^m$};
        \node[below right] at (B) {$\bfphi_{j}^m$};
        \node[left] at (Bnorm) {$\bfn_j^m$};
        \node[left] at (C) {$\bfphi_{j-1}^m$};
        
        \draw[very thick,->,blue] (C) -- (A) node[midway,below,rotate=-22.5] {$\bfphi_{j+1}^m-\bfphi_{j-1}^m$};
        \draw[dashed] (Bmean) -- (B);
        \draw[very thick,->] (B) -- (Bnorm);
        \node at (A)[circle,fill,inner sep=1.5pt]{};
        \node at (B)[circle,fill,inner sep=1.5pt]{};
        \node at (C)[circle,fill,inner sep=1.5pt]{};
    \end{tikzpicture}
    \caption{An illustration of the numerical approximation procedure for finding the approximate normal at node $\bfphi^m_j$. The displacement vector between nodes adjacent nodes (indicated in blue) is normalised and rotated anticlockwise, giving $\bfn_j^m$.}
    \label{fig:Normal}
\end{figure}

\smallskip\noindent\textbf{Collision detection.}
An important feature of the Frank--Read source is that dislocation lines eventually come into contact, and we must handle these collisions. When a collision occurs, segments with the same Burgers vector and opposite orientation annihilate, resulting in a topological change in which two new dislocation lines are formed, one segment being fixed at the pinning points, and another forming a closed loop that continues to expand. To ensure this phenomenon is accurately described we must calculate whether an intersection occurs during a time step, and if so, exactly where and when the collision happens. To do this, we adopt a collision detection algorithm called interval halving (which is often used for collision detection in video games). An implementation of this algorithm is described in \cite{Ericson05}.

After each time step, we calculate the minimum distance between the segments and the maximum distance each segment moves over the course of the time step: this is bounded above by the maximum distance the end points of the segment move. We can check if a collision has occurred in the time interval by comparing the minimum separation distance between the two segments before and after the time step has occurred. A collision has occurred only if the minimum separation distances before and after the time step must is less than the maximal distance travelled by the segments during the time step. If a collision is detected, the time interval is split in half and each of the resulting time subintervals are checked for a collision, proceeding until the collision time is determined to within a small tolerance, $\epsilon_\text{interval}$.

This procedure is repeated for all segments and if a collision has been detected we take the earliest collision time out of all detected collisions as the final collision time. The intuitive idea behind the algorithm is that we binary search the movement of two segments between a time step for collisions. 

A naive implementation of the algorithm described above would have $O(N^2)$ scaling with $N$ being the number of segments. To improve computational efficiency, we only need compare segments which are sufficiently close to each other. More specifically, in the simulations conducted, collisions always occur below and between the two pinned points. This area of interest is divided into a grid, where each square has side length equal to $\delta_\text{max}$. We then identify which square cell each point lies in, which can be done in $O(N)$ time. When comparing segments we go cell by cell, taking one segment in the middle cell and only comparing it to the segments which are in cells directly adjacent or diagonal to the current cell.

\smallskip\noindent\textbf{Topological cutting.}
If a collision is detected, the dislocation loop is split into two by removing the segments which have collided. 

After this step there will be a new curve connecting the two pinning points and a separate closed loop, and we can then continue the evolution by time-stepping as before, until the next collision is detected.

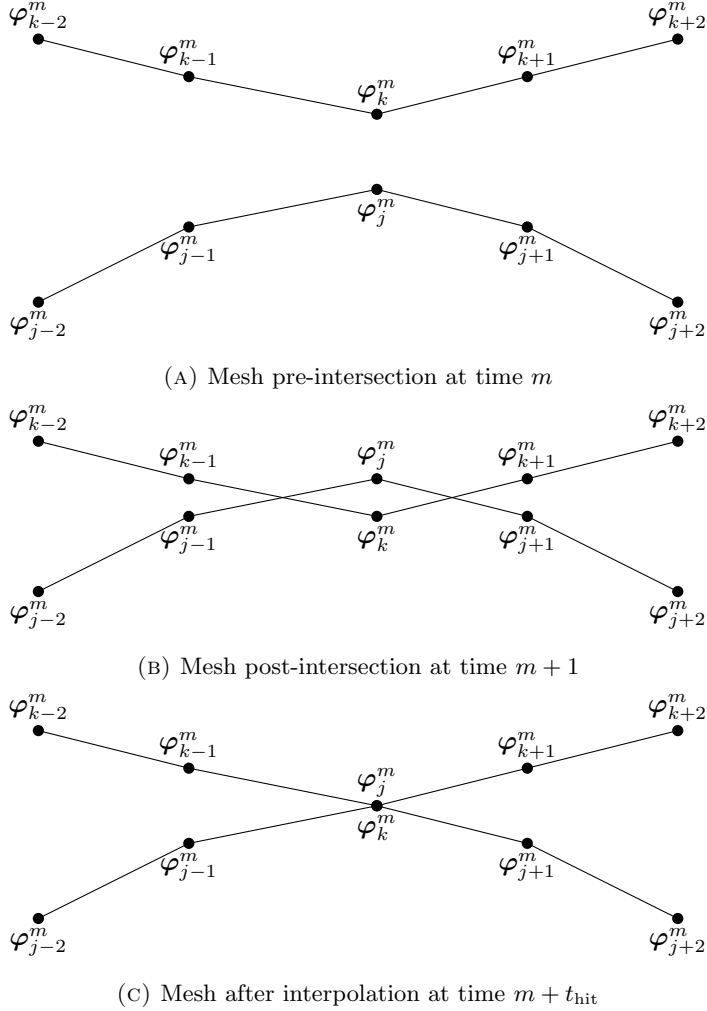
\begin{figure}[tbp]
\begin{subfigure}[b]{1\linewidth}
        \centering
    \begin{tikzpicture}
    \tikzset{>=latex}
        \tkzDefPoint(4,-1.5){Z1};
        \tkzDefPoint(2,-0.5){A1};
        \tkzDefPoint(0,0){B1};
        \tkzDefPoint(-2.5,-0.5){C1};
        \tkzDefPoint(-4.5,-1.5){D1};
        \tkzDefPoint(4,2){Z2};
        \tkzDefPoint(2,1.5){A2};
        \tkzDefPoint(0,1){B2};
        \tkzDefPoint(-2.5,1.5){C2};
        \tkzDefPoint(-4.5,2){D2};
        \tkzLabelPoint[below](Z1){$\bfphi_{j+2}^m$};
        \tkzLabelPoint[below](A1){$\bfphi_{j+1}^m$};
        \tkzLabelPoint[below](B1){$\bfphi_{j}^m$};
        \tkzLabelPoint[below](C1){$\bfphi_{j-1}^m$};
        \tkzLabelPoint[below](D1){$\bfphi_{j-2}^m$};
        \tkzLabelPoint[above](D2){$\bfphi_{k-2}^m$};
        \tkzLabelPoint[above](C2){$\bfphi_{k-1}^m$};
        \tkzLabelPoint[above](B2){$\bfphi_{k}^m$};
        \tkzLabelPoint[above](A2){$\bfphi_{k+1}^m$};
        \tkzLabelPoint[above](Z2){$\bfphi_{k+2}^m$};
        \node at (Z1)[circle,fill,inner sep=1.5pt]{};
        \node at (A1)[circle,fill,inner sep=1.5pt]{};
        \node at (B1)[circle,fill,inner sep=1.5pt]{};
        \node at (C1)[circle,fill,inner sep=1.5pt]{};
        \node at (D1)[circle,fill,inner sep=1.5pt]{};
        \node at (Z2)[circle,fill,inner sep=1.5pt]{};
        \node at (A2)[circle,fill,inner sep=1.5pt]{};
        \node at (B2)[circle,fill,inner sep=1.5pt]{};
        \node at (C2)[circle,fill,inner sep=1.5pt]{};
        \node at (D2)[circle,fill,inner sep=1.5pt]{};
        \draw[] (Z1) -- (A1) -- (B1) -- (C1) -- (D1);
        \draw[] (Z2) -- (A2) -- (B2) -- (C2) -- (D2);
    \end{tikzpicture}
    \caption{Mesh pre-intersection at time $m$}
\end{subfigure}
\begin{subfigure}[b]{1\linewidth}
        \centering
    \begin{tikzpicture}
    \tikzset{>=latex}
        \tkzDefPoint(4,-1){Z1};
        \tkzDefPoint(2,0){A1};
        \tkzDefPoint(0,0.5){B1};
        \tkzDefPoint(-2.5,0){C1};
        \tkzDefPoint(-4.5,-1){D1};
        \tkzDefPoint(4,1){Z2};
        \tkzDefPoint(2,0.5){A2};
        \tkzDefPoint(0,0){B2};
        \tkzDefPoint(-2.5,0.5){C2};
        \tkzDefPoint(-4.5,1){D2};
        \tkzLabelPoint[below](Z1){$\bfphi_{j+2}^m$};
        \tkzLabelPoint[below](A1){$\bfphi_{j+1}^m$};
        \tkzLabelPoint[above](B1){$\bfphi_{j}^m$};
        \tkzLabelPoint[below](C1){$\bfphi_{j-1}^m$};
        \tkzLabelPoint[below](D1){$\bfphi_{j-2}^m$};
        \tkzLabelPoint[above](D2){$\bfphi_{k-2}^m$};
        \tkzLabelPoint[above](C2){$\bfphi_{k-1}^m$};
        \tkzLabelPoint[below](B2){$\bfphi_{k}^m$};
        \tkzLabelPoint[above](A2){$\bfphi_{k+1}^m$};
        \tkzLabelPoint[above](Z2){$\bfphi_{k+2}^m$};
        \node at (Z1)[circle,fill,inner sep=1.5pt]{};
        \node at (A1)[circle,fill,inner sep=1.5pt]{};
        \node at (B1)[circle,fill,inner sep=1.5pt]{};
        \node at (C1)[circle,fill,inner sep=1.5pt]{};
        \node at (D1)[circle,fill,inner sep=1.5pt]{};
        \node at (Z2)[circle,fill,inner sep=1.5pt]{};
        \node at (A2)[circle,fill,inner sep=1.5pt]{};
        \node at (B2)[circle,fill,inner sep=1.5pt]{};
        \node at (C2)[circle,fill,inner sep=1.5pt]{};
        \node at (D2)[circle,fill,inner sep=1.5pt]{};
        \draw[] (Z1) -- (A1) -- (B1) -- (C1) -- (D1);
        \draw[] (Z2) -- (A2) -- (B2) -- (C2) -- (D2);
    \end{tikzpicture}
    \caption{Mesh post-intersection at time $m+1$}
\end{subfigure}
\begin{subfigure}[b]{1\linewidth}
        \centering
    \begin{tikzpicture}
    \tikzset{>=latex}
        \tkzDefPoint(4,-1.5){Z1};
        \tkzDefPoint(2,-0.5){A1};
        \tkzDefPoint(0,0){B1};
        \tkzDefPoint(-2.5,-0.5){C1};
        \tkzDefPoint(-4.5,-1.5){D1};
        \tkzDefPoint(4,1){Z2};
        \tkzDefPoint(2,0.5){A2};
        \tkzDefPoint(0,0){B2};
        \tkzDefPoint(-2.5,0.5){C2};
        \tkzDefPoint(-4.5,1){D2};
        \tkzLabelPoint[below](Z1){$\bfphi_{j+2}^m$};
        \tkzLabelPoint[below](A1){$\bfphi_{j+1}^m$};
        \tkzLabelPoint[above](B1){$\bfphi_{j}^m$};
        \tkzLabelPoint[below](C1){$\bfphi_{j-1}^m$};
        \tkzLabelPoint[below](D1){$\bfphi_{j-2}^m$};
        \tkzLabelPoint[above](D2){$\bfphi_{k-2}^m$};
        \tkzLabelPoint[above](C2){$\bfphi_{k-1}^m$};
        \tkzLabelPoint[below](B2){$\bfphi_{k}^m$};
        \tkzLabelPoint[above](A2){$\bfphi_{k+1}^m$};
        \tkzLabelPoint[above](Z2){$\bfphi_{k+2}^m$};
        \node at (Z1)[circle,fill,inner sep=1.5pt]{};
        \node at (A1)[circle,fill,inner sep=1.5pt]{};
        \node at (B1)[circle,fill,inner sep=1.5pt]{};
        \node at (C1)[circle,fill,inner sep=1.5pt]{};
        \node at (D1)[circle,fill,inner sep=1.5pt]{};
        \node at (Z2)[circle,fill,inner sep=1.5pt]{};
        \node at (A2)[circle,fill,inner sep=1.5pt]{};
        \node at (B2)[circle,fill,inner sep=1.5pt]{};
        \node at (C2)[circle,fill,inner sep=1.5pt]{};
        \node at (D2)[circle,fill,inner sep=1.5pt]{};
        \draw[] (Z1) -- (A1) -- (B1) -- (C1) -- (D1);
        \draw[] (Z2) -- (A2) -- (B2) -- (C2) -- (D2);
        
    \end{tikzpicture}
    \caption{Mesh after interpolation at time $m+ t_\text{hit}$}
\end{subfigure}

    \caption{ 
    Illustration of merging, the interpolated mesh at time $m+t_\text{hit}$ replaces the mesh at time $m+1$ and the simulation continues with the split, interpolated mesh.}
    \label{fig:enter-label}
\end{figure}

\smallskip\noindent\textbf{Re-meshing.}
As the Frank--Read source evolution equation \eqref{eq:FR_parametric} contains a forcing term, this generally causes the internodal lengths to grow over time, in contrast with the evolution under MCF. To maintain accuracy, we must therefore frequently both add new nodes by linearly interpolating between nodes that are too far away from each other, and remove nodes that are too close together. Changing the number of nodes at each time step causes $N$ to depend on $m$, but since we opt for a single-step explicit approach to the time evolution, one can see that this does not pose an issue for the scheme as described in \eqref{lumped_fully_discrete_FE_MCF}. 
The approach to remeshing we choose to adopt is based purely on the relative separation between adjacent nodes. To save computational cost, more sophisticated remeshing schemes based upon \emph{a posteriori} error estimates could likely be devised depending upon the position and angle between adjacent nodes, but are beyond the scope of the present work. 

To describe the approach we use, we introduce parameters $\delta_\text{min}$ and $\delta_\text{max}$, which are the minimum and maximum (dimensionless) distances allowed between vertices. Let $\Phi$ be the set that contains the re-meshed points of $\set{\bfphi^m_j}_{j=1}^{N}$. Initially we set $\Phi = \set{\bfphi^m_1}$. We then consider the internodal distance $q^m_2:=|\bfphi^m_2-\bfphi^m_1|$ and proceed in one of three ways:
\begin{enumerate}
\item If $q^m_2 \in [\delta_\text{min},  \delta_\text{max}]$ then we add $\bfphi^m_2$ to $\Phi$.
\item If $q^m_2 > \delta_\text{max}$, then we add the midpoint between $\bfphi^m_1$ and $\bfphi^m_2$ to $\Phi$, i.e. we add the vector $\frac 1 2 \brac{\bfphi^m_2-\bfphi^m_1}$ to $\Phi$. We then add $\bfphi^m_2$ to $\Phi$.
\item If  $q^m_2 < \delta_\text{min}$, then we find the first $2<k < N$ such that $\sum^k_{i=2}q_i^m >\delta_\text{min}$ and then add $q^m_k$ to $\Phi$. If no such $k$ exists then we continue without adding anything to $\Phi$.
\end{enumerate} 
Let $\bfphi^m_j$ be the last node added to $\Phi$. We then repeat these three steps with $q^m_{j+1}$ until we add $q^m_{N-1}$. If the curve is closed, when considering $\bfphi_N^m$ we must check that both $q^m_N$ and $q^m_1$ are both greater than $\delta_\text{min}$. If $q^m_N$ or $q^m_1$ is greater than $\delta_\text{max}$ we must also add interpolated nodes. We then start the whole process again and keep repeating with the newly calculated $q^m_j$, until we find that for all $i \in \set{1, \dotsc, N}$ we have that $q^m_i \in [\delta_\text{min},  \delta_\text{max}]$ (i.e. in the last pass we never take option 2 or 3).

\smallskip\noindent\textbf{Summary.}
All of the algorithms discussed above are summarised in pseudocode in Algorithm~\ref{MCF_Top_Change_Alg}.

In practice, the algorithms described were implemented in C\# as a .NET project, which can be accessed at~\cite{Rydell24frsmcfrepo}.

In the code there are five classes/structs:
\begin{itemize}
    \item \texttt{Program}, which is contains the main method that is called first when the program starts to run and methods $x$ and $y$ which are used to parameterize $\bfphi^0(\theta) = (x(\theta), y(\theta))$.
    \item \texttt{Flow},  a class that contains all the information and methods relevant to a dislocation line's evolution through time, as well as methods for checking for intersections and implementing topological change etc.
    \item \texttt{MCF}, a class that contains the relevant methods for calculating the movement of a Flow object by MCF.
    \item \texttt{fPoint}, a struct that stores 2D vectors and methods for manipulating them.
    \item \texttt{FlowReadWrite}, a class that contains methods for writing the data of the simulation to pictures and videos.
\end{itemize}

We found that choosing the values $\delta_\text{max} = 5 \times 10^{-3}$ and $\delta_\text{min} =5 \times 10^{-4}$ provided an appropriate compromise between computational intensity while maintaining accuracy of a dislocation line's shape as it evolves, and all simulations were performed using these values. Linear algebra operations required to solve the linear systems in order to time step were implemented ab initio.

\begin{table}[tbp]
    \centering
    \begin{tabular}{||c|c||}
    \hline
    Variable & Value \\
    \hline
      $\delta_\text{max}$ & $5 \times 10^{-3}$  \\
       $\delta_\text{min}$  & $5 \times 10^{-4}$ \\  
       $\epsilon_\text{interval}$ & $1 \times 10^{-7}$ \\
       \hline
    \end{tabular}
    \caption{Default tolerance parameters used.}
    \label{tab:tolerances}
\end{table}

\section{Results}
\label{sec:results}
We now present results obtained from the computational model described above. As we will see, even though our model does not take into account any long-range elastic interaction effects, we are able to reproduce experimental pictures with surprising accuracy. We also use the model to make predictions about the rate at which dislocation line length is produced by such a source which could be tested experimentally.

\subsection{Validation against experiment}
As discussed in the introduction, experimental observations of approximately isotropic Frank--Read sources are relatively few. Nevertheless, Figure~\ref{fig:experiment} provides evidence that our model is valid for some experimentally-observed physical systems; in this case, the picture is taken from a Ni-Fe alloy. Although this figure provides no indication of scale, this provides no barrier to comparison: Our dimensionless model enables a simple parametric search allowing us to obtain excellent qualitative agreement with the experimental picture. In particular, the relative sizes of all observed dislocation line pieces are accurately resolved, even including small details such as the dent at the bottom of the just formed outermost loop.

\subsection{General features of the evolution}
\begin{figure}[tbp] 
  \centering
\begin{subfigure}{.3\textwidth}

  \includegraphics[width=\linewidth]{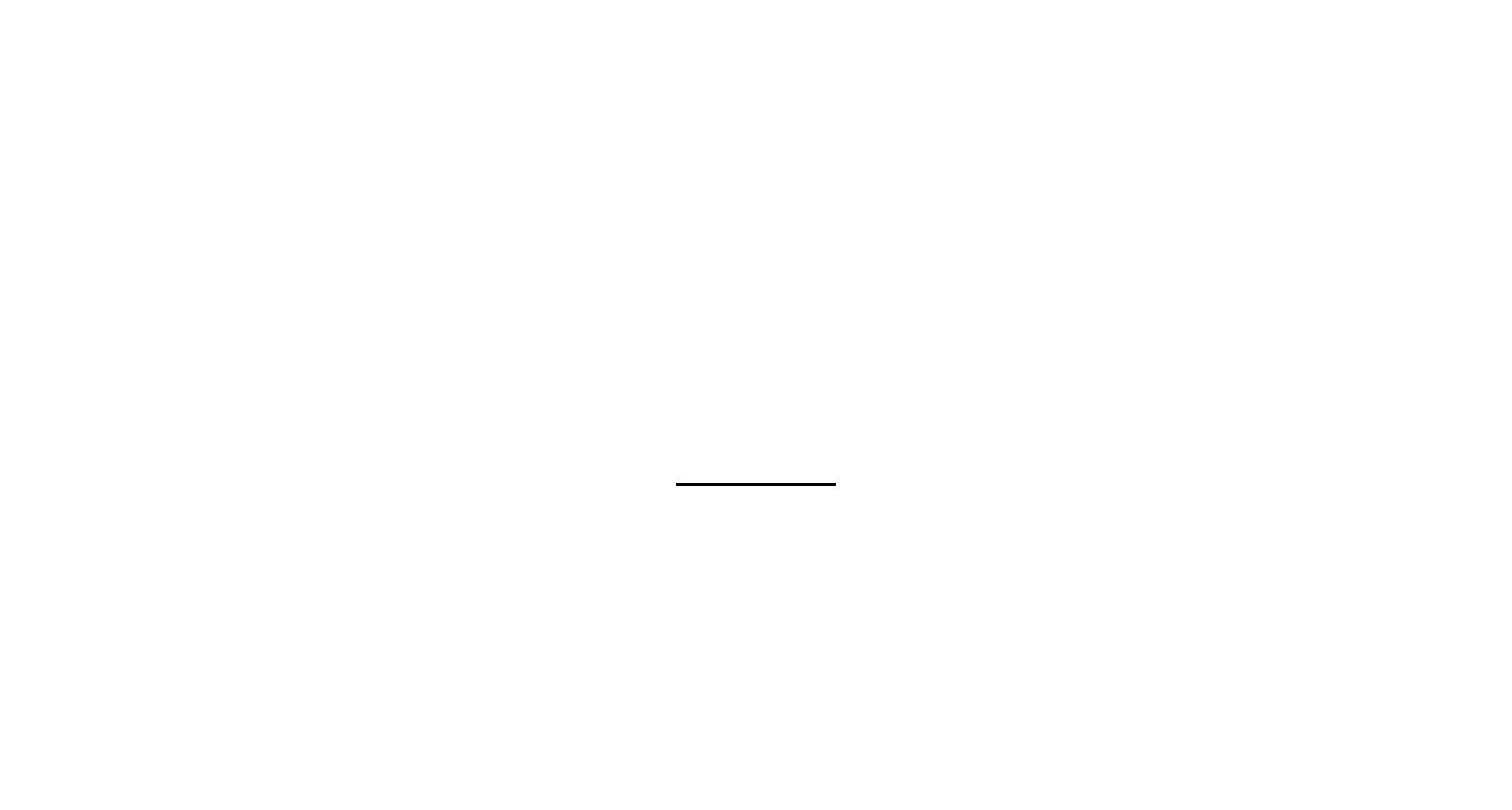}
  \caption{$t=0.00$} 
\end{subfigure}
\begin{subfigure}{.3\textwidth}
  \includegraphics[width=\linewidth]{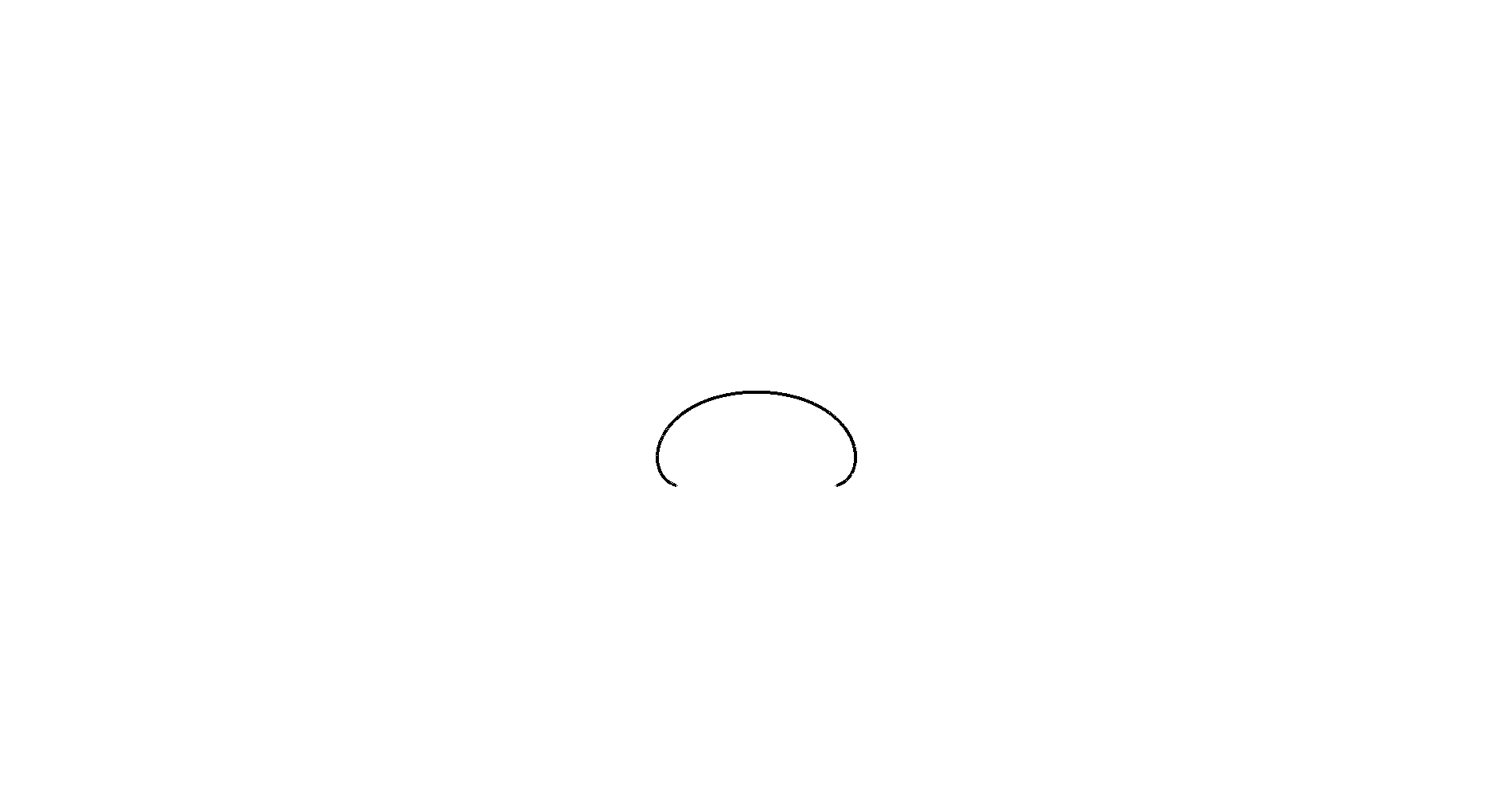}
  \caption{$t=0.30$} 
\end{subfigure}
\begin{subfigure}{.3\textwidth}
  \includegraphics[width=\linewidth]{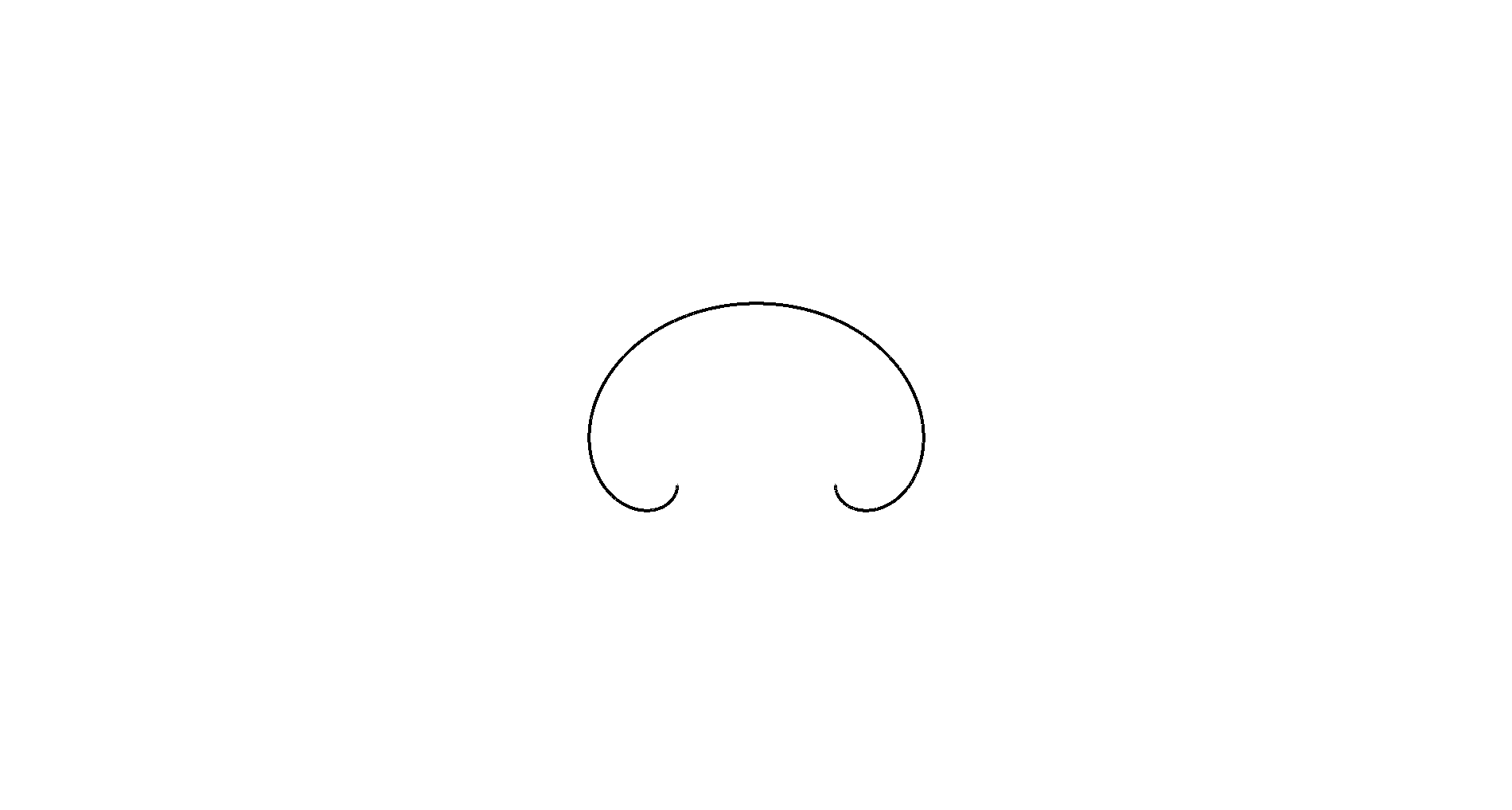}
  \caption{$t=0.60$} 
\end{subfigure}
\begin{subfigure}{.3\textwidth}
  \includegraphics[width=\linewidth]{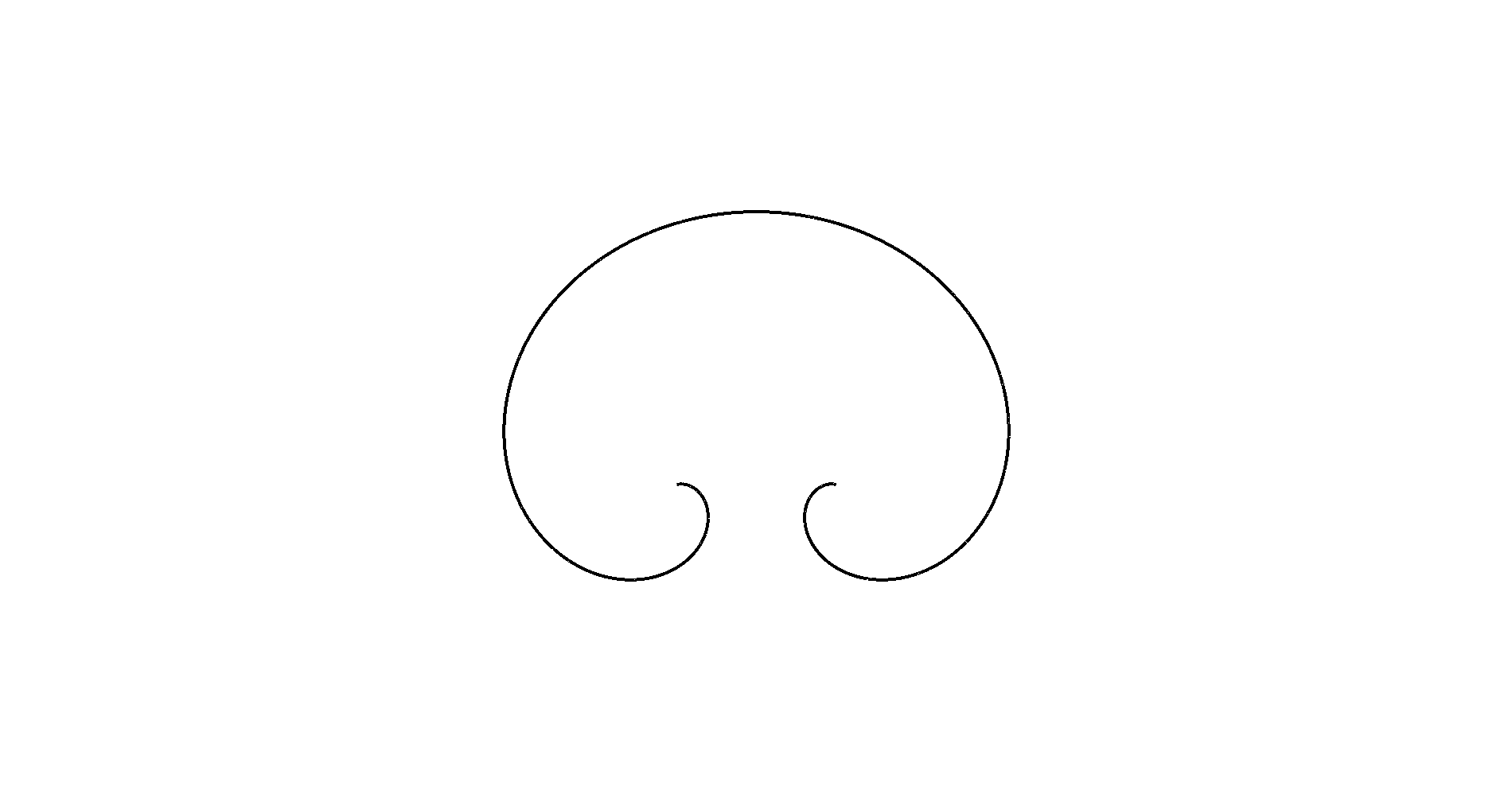}
  \caption{$t=0.90$} 
\end{subfigure}
\begin{subfigure}{.3\textwidth}
  \includegraphics[width=\linewidth]{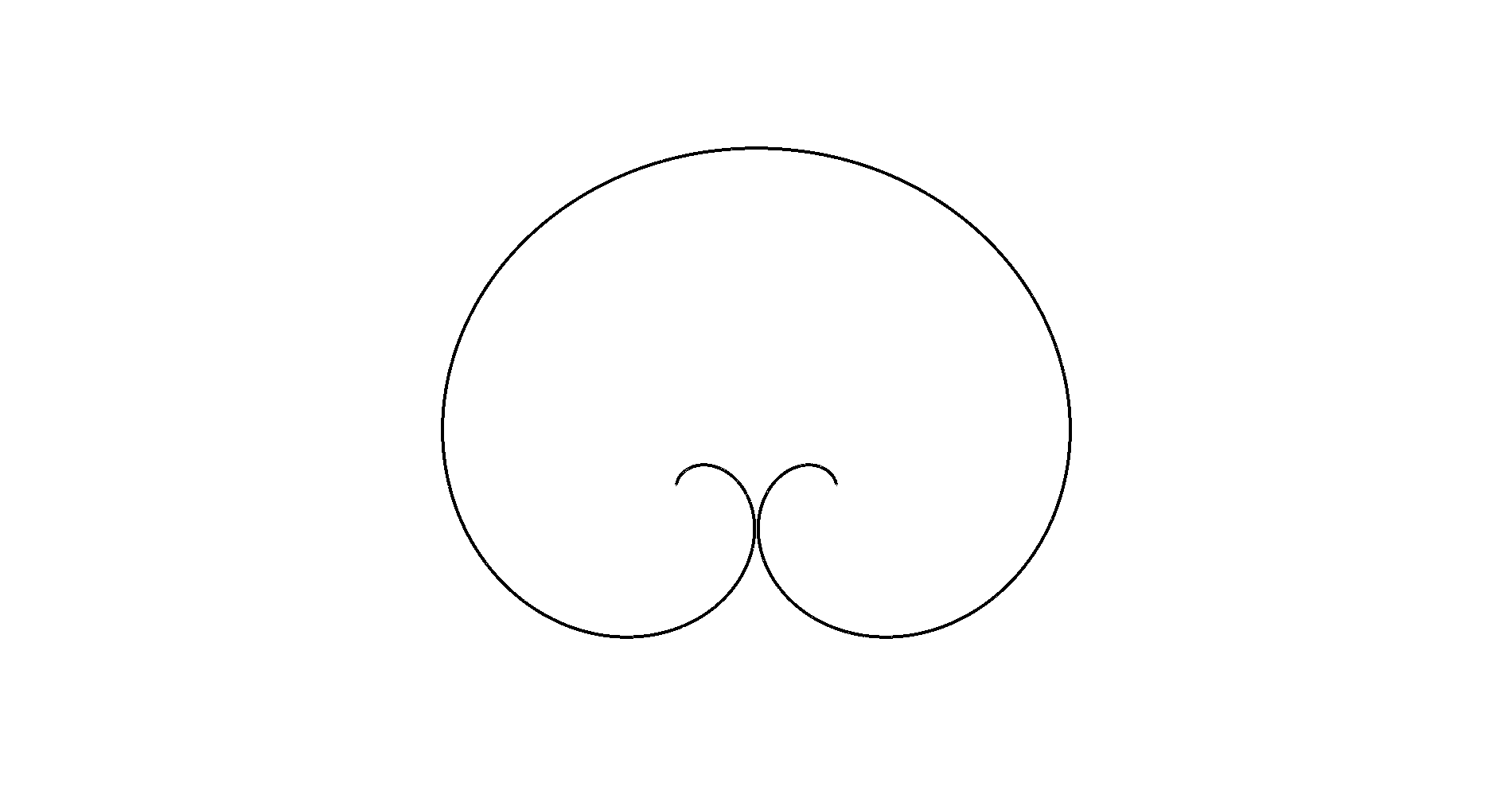}
  \caption{$t=1.10$} 
\end{subfigure}
\begin{subfigure}{.3\textwidth}
  \includegraphics[width=\linewidth]{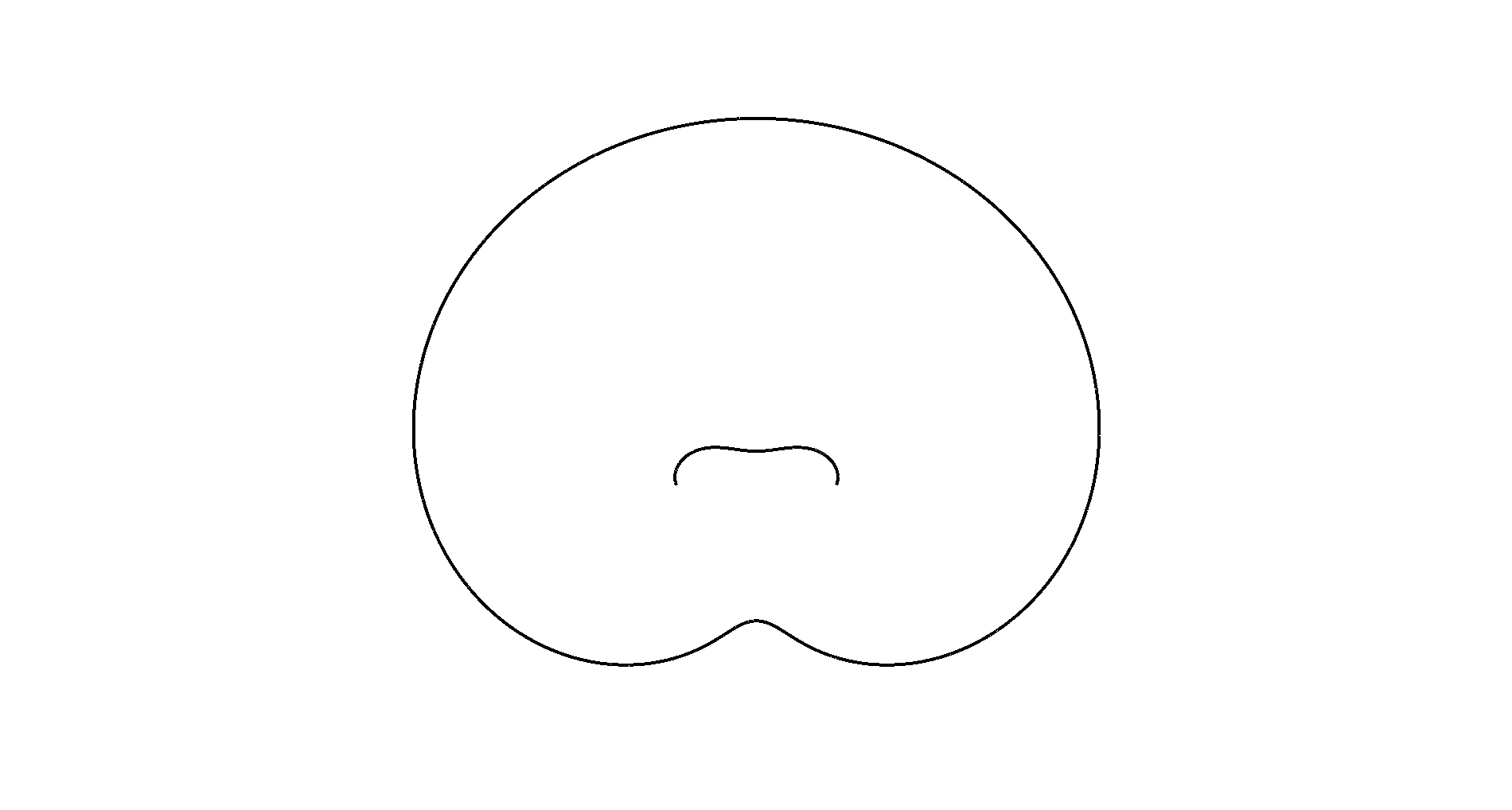}
  \caption{$t=1.20$} 
\end{subfigure}
\caption{Early-time snapshots of the evolution with $f_s = 5 \times 10^{-3}$. Simulation time step $\tau = 1.2 \times 10^{-3}$.}
\label{fig:time_evolution}
\end{figure}

\begin{figure}[tbp]
    \centering
    \includegraphics[width=0.5\linewidth]{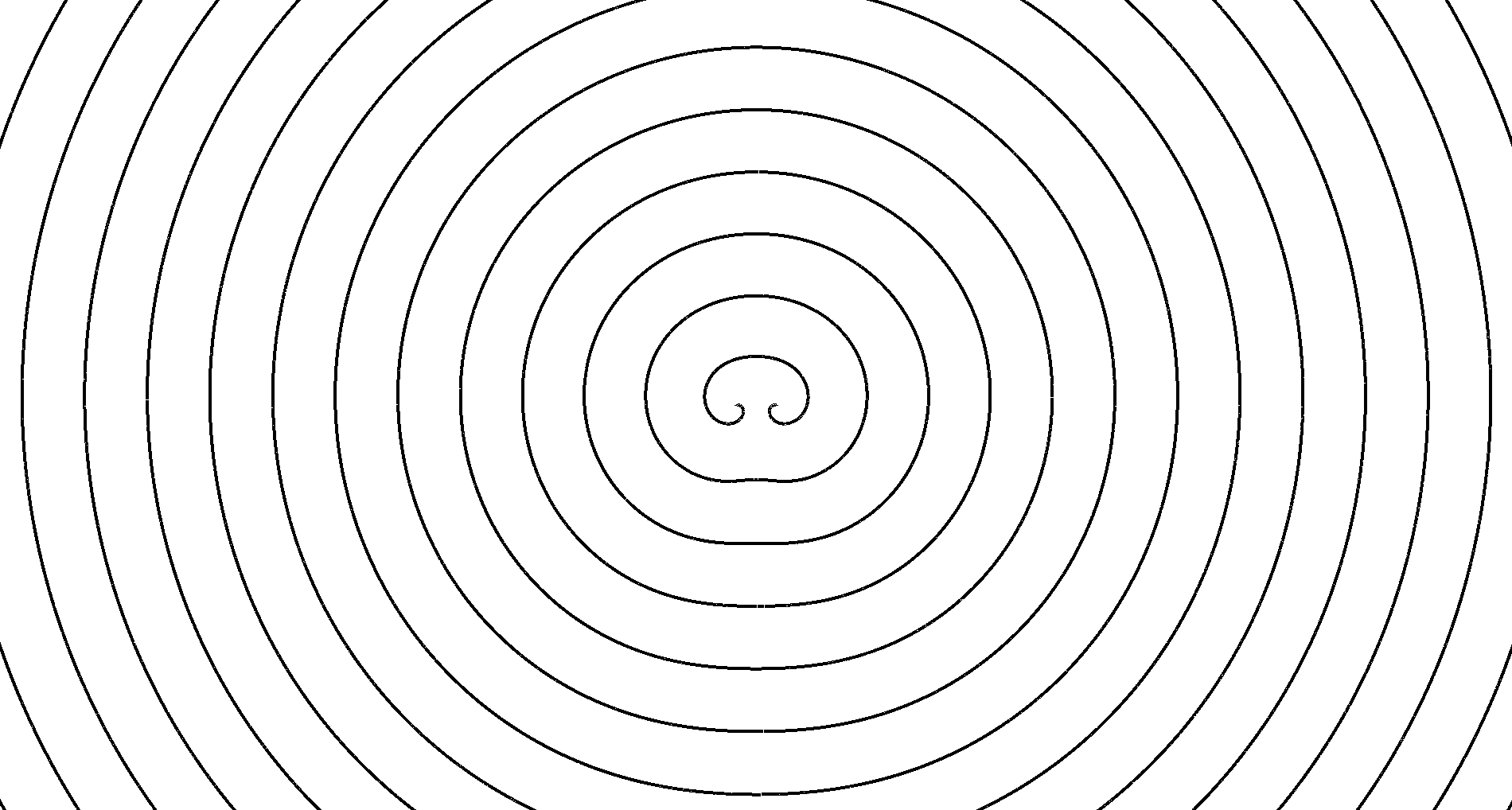}
    \caption{A late-time picture of the source at time $t = 9.18$. The parameter $f_s = 7 \times 10^{-3}$, and simulation time step is $\tau = 1.2 \times 10^{-3}$.}
    \label{fig:many_loops}
\end{figure}

Figure~\ref{fig:time_evolution} shows an indicative example trajectory of the Frank--Read source at early times, and Figure~\ref{fig:many_loops} shows a picture after a series of collisions have occured, showing a series of expanding loops which become more circular further from the pinning points. In all cases, the initial condition is taken to be a straight line connecting the pinning points.

\begin{figure}[tbp] 
  \centering
\begin{subfigure}{.3\textwidth}
  \includegraphics[width=\linewidth]{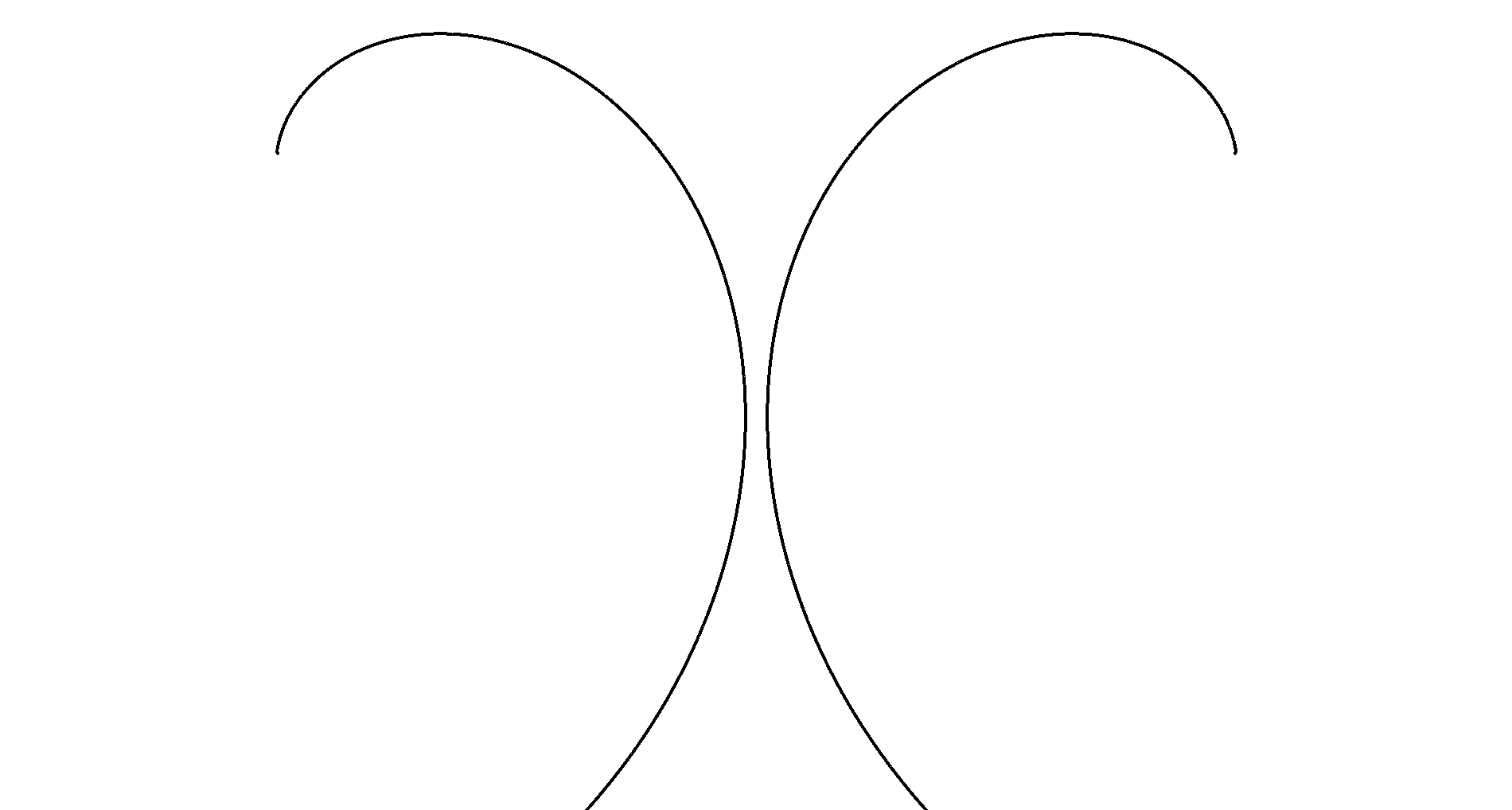}
  \caption{$t=1.104$} 
\end{subfigure}
\begin{subfigure}{.3\textwidth}
  \includegraphics[width=\linewidth]{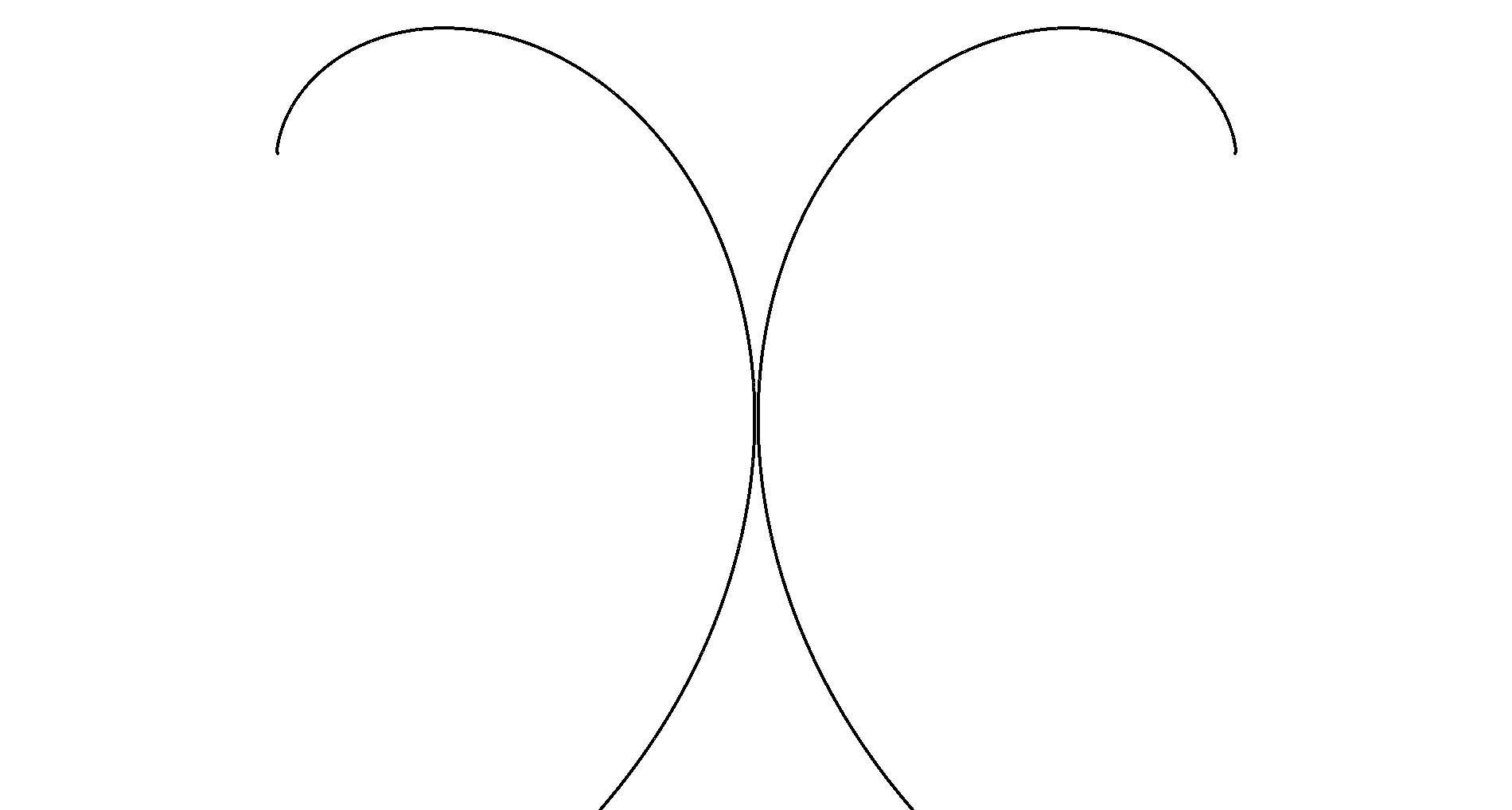}
  \caption{$t=1.110$} 
\end{subfigure}
\begin{subfigure}{.3\textwidth}
  \includegraphics[width=\linewidth]{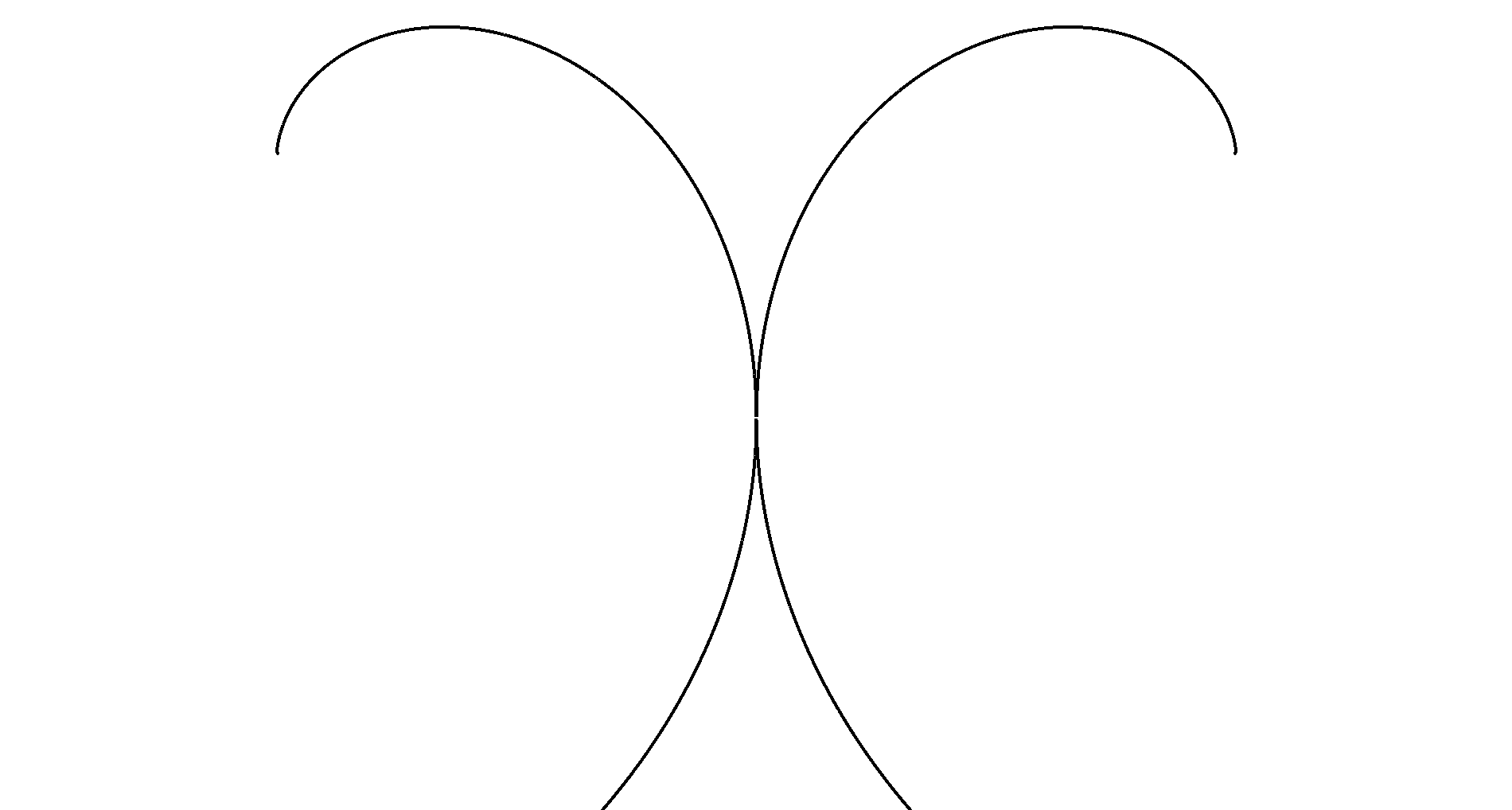}
  \caption{$t=1.111$} 
\end{subfigure}\\
\begin{subfigure}{.3\textwidth}
  \includegraphics[width=\linewidth]{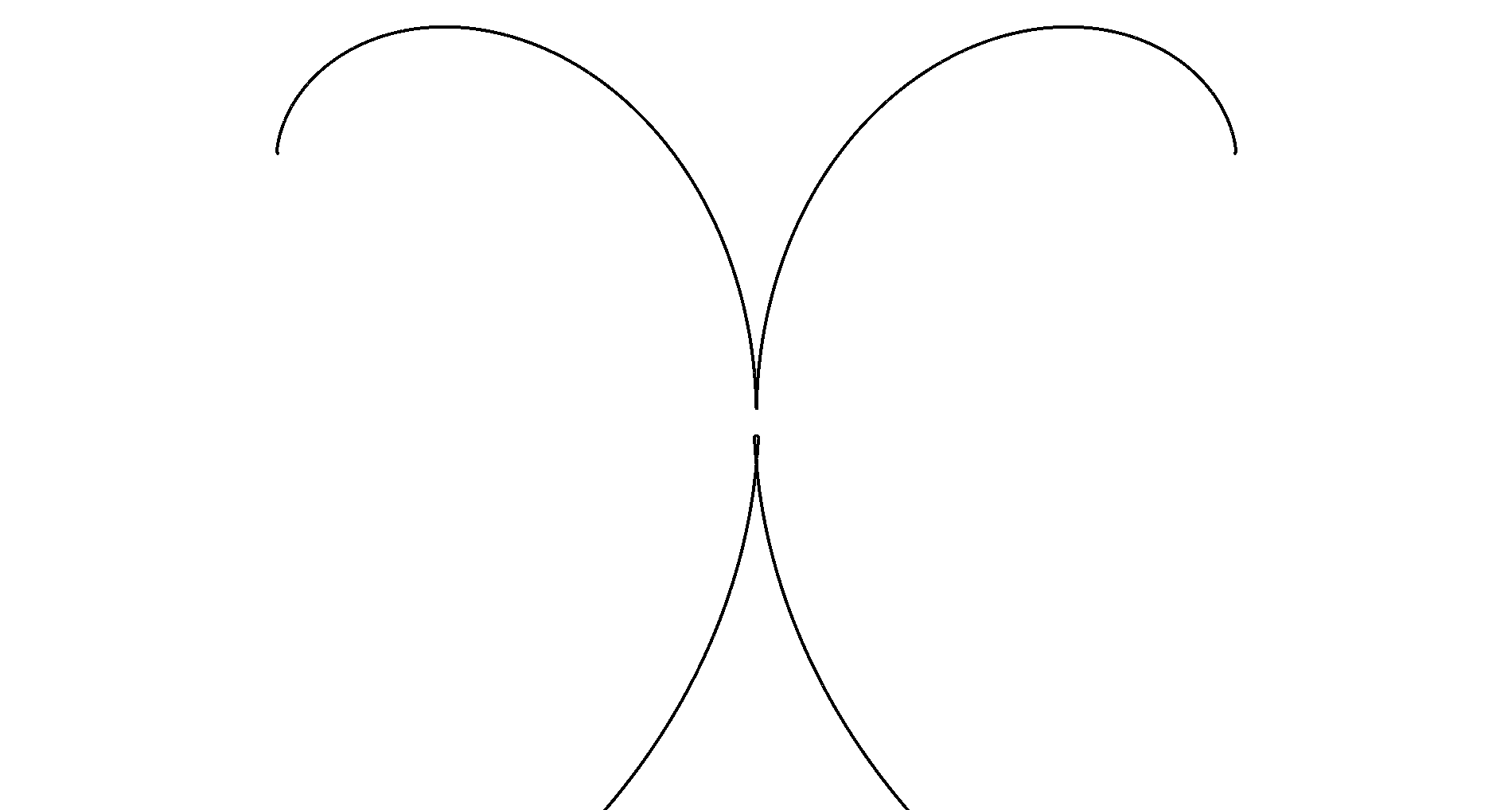}
  \caption{$t=1.112$} 
\end{subfigure}
\begin{subfigure}{.3\textwidth}
  \includegraphics[width=\linewidth]{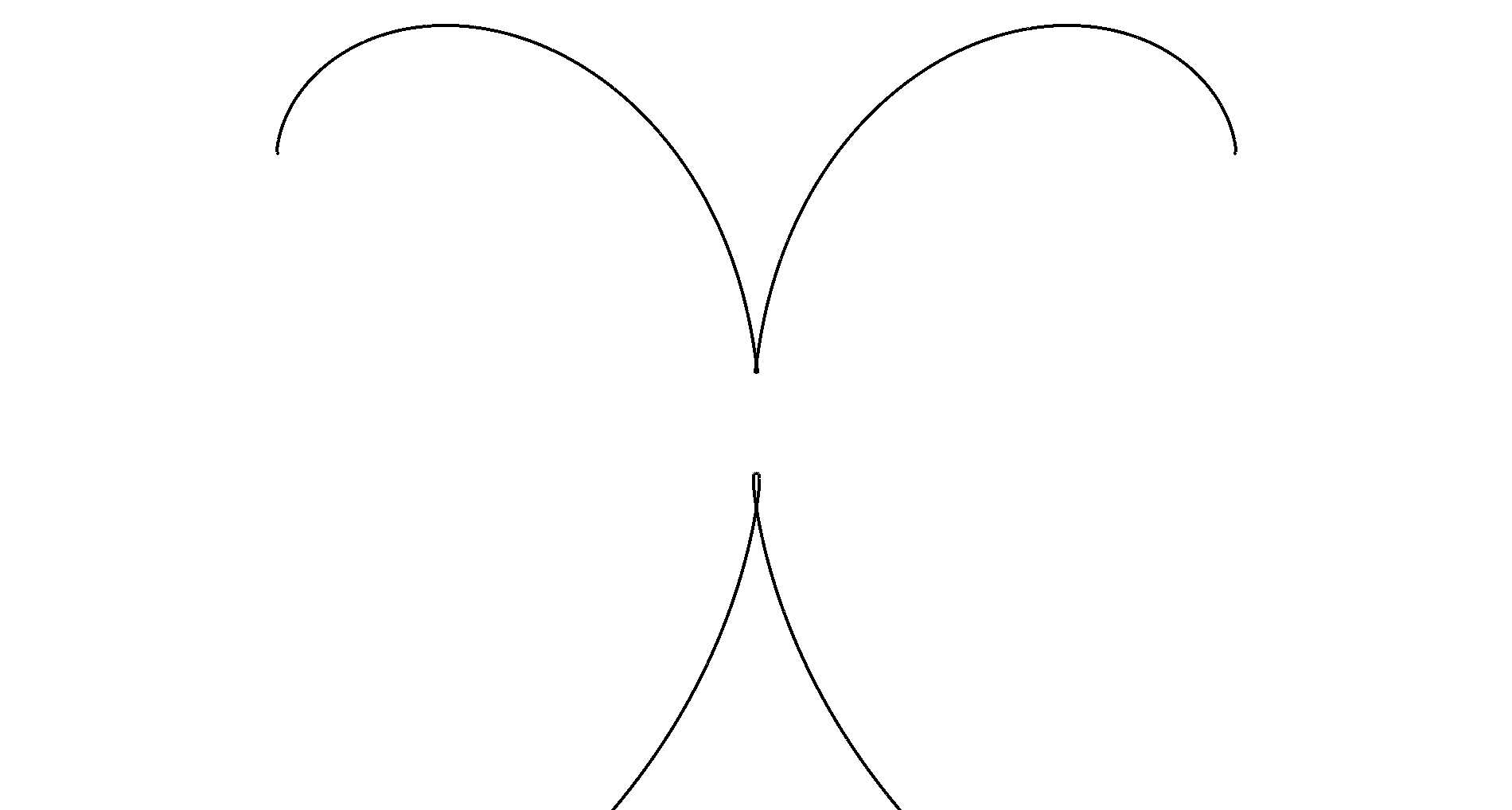}
  \caption{$t=1.116$} 
\end{subfigure}
\begin{subfigure}{.3\textwidth}
  \includegraphics[width=\linewidth]{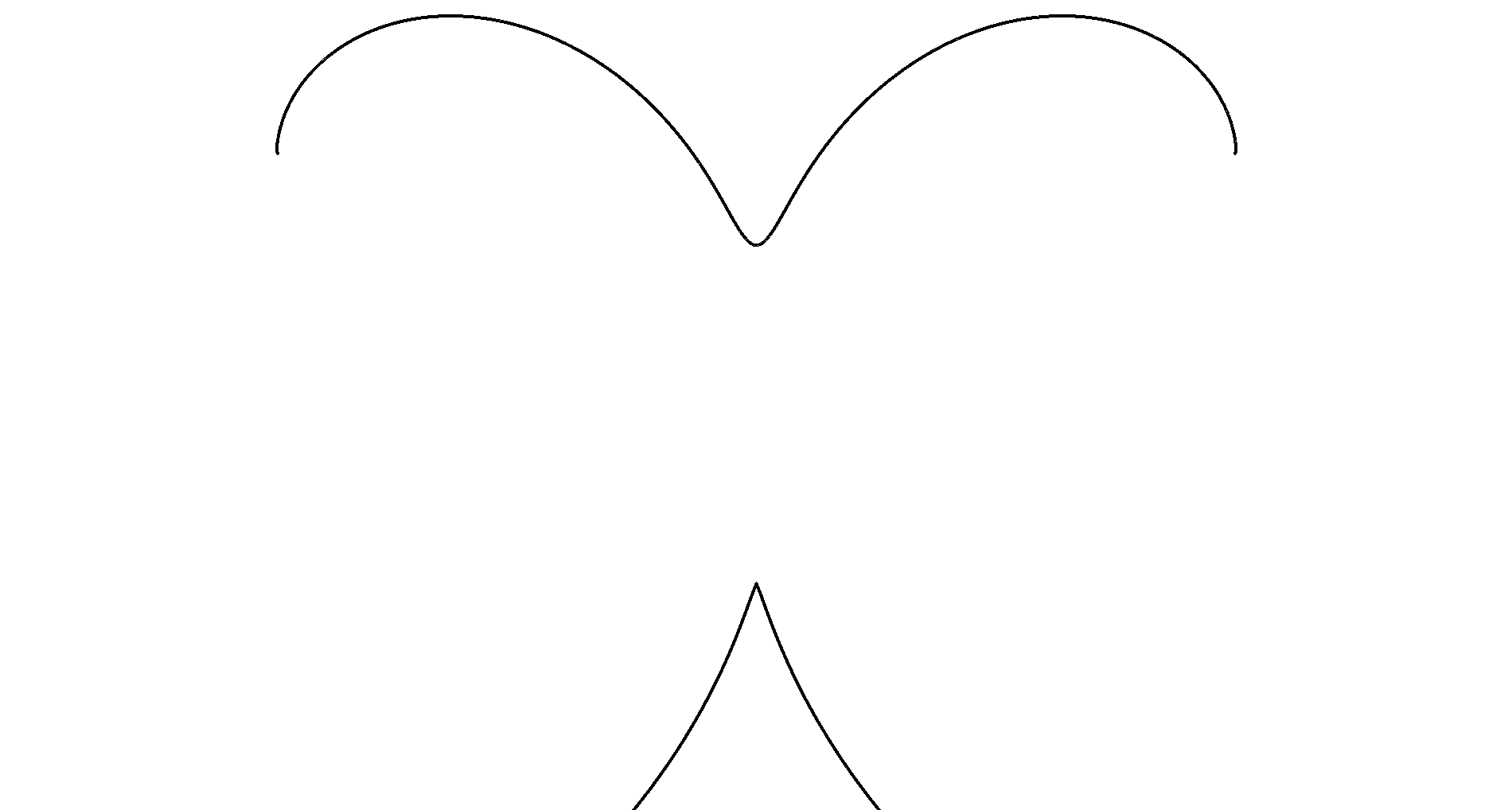}
  \caption{$t=1.126$} 
\end{subfigure}
\caption{Magnified snapshots of topological change occurring; (c) is show the simulation time where this occurs.
Here, $f_s = 5 \times 10^{-3}$ and the time step was $\tau = 1.2 \times 10^{-3}$.}
\label{fig:collision}
\end{figure}

The collision region is shown magnified at times close to the first collision in Figure~\ref{fig:collision}. The cusps generated by the collision rapidly move away from each other both due to the resulting high curvature and external driving force. At later times, further collisions occur in the same general area of the plane.

\subsection{Dislocation length growth over time}

As we have seen above, new closed loops are generated at a steady rate by the Frank--Read source, and a plot of the resulting length growth was shown in Figure~\ref{fig:length}. This indicates that at late times, the length grows at an approximately quadratic rate, and Figure~\ref{fig:many_loops} indicates that much of the dislocation length is stored in approximately circular loops which move outwards at a steady rate. To explain and analyse this rate, we propose a simplified model below.

\smallskip\noindent\textbf{Analysing length growth.}
Consider a circular loop $\gamma$ of radius $\rho$, each element of which is moving outwards with a uniform speed $c$ so that the velocity at any point on the curve is $c\bfn$. Over the period from time $t$ to $t+\delta t$, a point on the circular curve $\rho(\cos\theta,\sin\theta)$ moves to $(\rho+c\,\delta t)(\cos\theta,\sin\theta)$, so the new length of the curve is $2\pi(\rho+c\delta t)$. It follows that the rate of change of the curve length is
\[
\lim_{\delta t\to 0}\frac{2\pi(\rho+c\delta t)-2\pi\rho}{\delta t} = 2\pi c.
\]
As a result, the instantaneous change in the length of this single loop is
\[
\frac{\dd}{\dt}\text{Length}(\gamma(t)) = 2\pi c.
\]

We assume that new circular loops are created at an approximately constant rate $r$ (generations per time unit). Denote by $\gamma_0,\gamma_1,\gamma_2,\ldots$ the loops generated at times $0, r^{-1}, 2 r^{-1}, \ldots$. Thus, we have the following expression for the total length of dislocations at time $t$, up to which $N = \lfloor tr \rfloor$ loops have been generated:
\[
  \ell = \sum_{j = 0}^N \text{Length}(\gamma_j(t))
  = \sum_{j = 0}^N 2\pi c (t - jr^{-1})
\]
For $t$ large, we can approximate this sum via an integral as follows:
\begin{align*}
  \ell &= 2\pi c r \sum_{j=0}^{\lfloor tr \rfloor} (t - jr^{-1}) r^{-1} \\
  &\approx 2\pi c r \int_0^t t - s \,\ds  \\
  &= 2\pi c r \biggl[-\frac12(t-s)^2\biggr]_{s=0}^{t} \\
  &= \pi c r t^2.
\end{align*}
Combining all proportionality constants, this gives the law claimed in the introduction,
\[
  \ell \sim t^2.
\]
Note that this formula is only valid assuming that the generated loops are roughly circular and the discrete nature of loop generation is unimportant. Both assumptions are true for large times $t$ since then the more intricate shape of the newly created dislocations loops can be neglected and the oscillation frequency of the innermost loop in the Frank--Read source is sufficiently high relative to the time scale, so that we may consider the loop creation to be essentially continuous. The numerical evidence for this is presented in the next section.

\smallskip\noindent\textbf{Comparing with simulation data.}
A simulation was carried out for 36s ($3 \times 10^4$ time steps) with $f_s = 5.0\times 10^{-3}$ and a time step of $\tau  = 1.2\times 10^{-3}$. The first topological change occurs at time $1.11 $, and then new dislocation lines are produced consistently at a rate of every $1.07 $ dislocation lines per second. 

\smallskip\noindent\textbf{Ordinary least squares fitting of log--log plot.}
When fitting a curve to the log--log graph of the length plot, for later times we can ignore lower order terms to get 
\[
\log L(t) \approx \log(\pi r c) + 2\log t,
\]
and allowing us to to get a linear graph in $\log t$ with a slope of 2. Furthermore, at late times we can assume that forcing due to curvature is negligible compared to $f_s$ and therefore take $c \approx f_s = 5 \times 10^{-3}$. The total length of all dislocation lengths is calculated by summing the individual lengths of each segment. We then used the Python library \texttt{sci-kit} to apply ordinary least squares (OLS) to fit a line to the log--log graph of the length plot from times $18$ to $36$ time units ($1.8 \times 10^4$ to $3 \times 10^4$ time steps) against the values of $2\log(t)$. This produces a slope 1.994 and an intercept of 1.101 (both to three decimal places), suggesting
$$\log(\pi \tau r c) \approx 1.1.$$
This relationship may be rearranged to predict that the time to produce a new loop is approximately
$$r \approx 1.00 \,\text{s}.$$
The data produces a coefficient of determination $R^2 = 1.00$, suggesting that the log--log plot of $L(t)$ is essentially linear at later times. 

Furthermore, one can calculate the root mean squared error (RMSE), again using the $\texttt{sci-kit}$ library, of the true length values compared to the values predicted by OLS in the log-log plot. In this case we have the RMSE is $1.41 $. Table \ref{tab:regression_table} shows simulated and predicted $r$ values, along with the RMSE for different values of $f_s$. 

\begin{table}[tbp]
    \centering
    \begin{tabular}{||c||c||c||c||}
    \hline
    $f_s$ & Simulation  & OLS & RMSE\\
    \hline
       $3 \times 10^{-3}$ & 2.82   & 0.24 & 1.40 \\
       $5 \times 10^{-3}$ & 1.00 & $1.07$ & $1.41 $\\
       $7 \times 10^{-3}$ & 0.543  & 0.985  & 1.66 \\
       \hline
    \end{tabular}
    \caption{Simulated and predicted $r$ values along with root mean squared error. The simulation was run for $36$ time units ($3 \times 10^4$ time steps). The ordinary least squares regression was calculated by fitting the length data from between $18$ and $36$ time units ($ 1.8 \times 10^4$ to $3 \times 10^4$ time steps) against a quadratic rate in the log-log graph. All simulations used $\tau = 1.2 \times 10^{-3}$.}
    \label{tab:regression_table}
\end{table}

\section{Conclusions}
\label{sec:conclusion}
We have presented a derivation of a mathematical and computational model for a Frank--Read source for dislocations in crystalline materials. By neglecting long-range elastic interaction and considering dislocations in the line tension limit, the model obtained is a form of curve-shortening flow augmented by a forcing term. Through non-dimensionalisation, this was found to depend upon a single dimensionless number. The resulting simulation methodology was successfully validated against experimental pictures of Frank--Read sources. As a concrete outcome of the model, we predict the length growth of dislocations produced by the Frank--Read source over time, and the rate at which new dislocation loops are produced, both of which are quantities which could be investigated in greater detail experimentally.

Clearly, the validity of the model we have proposed here is limited to cases where long-range elastic interactions do not contribute in a significant way to the evolution, and as more dislocations are produced we can be less certain that this effect remains small. Furthermore, it is clear that anisotropy of the underlying crystal lattice may play an important role in the evolution, depending on the material system; compare for example the anisotropic loops observed in Si in \cite{Geipel1996}. Anisotropy of the line tension could be handled within a similar modelling framework, by introducing appropriate anisotropic variants of the forces described in Section~\ref{sec:derivation}, and this merits further investigation in future. Other new questions opened up by this work include a rigorous proof of the convergence of the scheme proposed, along with the development of a more sophisticated remeshing scheme based on \emph{a posteriori} error estimates to reduce the computational complexity when dislocation loops become large.

Finally, it would be interesting to validate the quantitative predictions of the present work against further experiments in specific materials. However, to implement this, precise experimental data is required, which we have not been able to locate.

\begin{algorithm}
\begin{algorithmic}
\caption{Move Nodes By Forced MCF}
\Function{MoveNodesByForcedMCF}{$\set{\bfphi^m_j}_{j=1}^N$}
\If{ $\set{\bfphi^m_j}_{j=1}^N$ is pinned}
\State Calculate $\bfphi^{m+1}_X$ by solving $A^m_\text{pinned}\bfphi^{m+1}_{X,\text{pinned}} =  B_{X,\text{pinned}}^m$
\State Calculate $\bfphi^{m+1}_Y$ by solving $A^m_\text{pinned}\bfphi^{m+1}_{Y,\text{pinned}} =  B_{Y,\text{pinned}}^m$
\ElsIf{$\set{\bfphi^m_j}_{j=1}^N$ is not pinned}
\State Calculate $\bfphi^{m+1}_X$ by solving $A^m\bfphi^{m+1}_X =  B_X^m$
\State Calculate $\bfphi^{m+1}_Y$ by solving $A^m\bfphi^{m+1}_Y =  B_Y^m$ 
\EndIf
\State Calculate  $\set{\bfn^{m+1}_j}_{j=1}^N$
\For{ $1 \leq j \leq N$}
\State $\bfphi_j^{m+1} \gets \bfphi_j^{m+1} + f_s\bfn^{m+1}_j  $
\EndFor
\State \Return $\set{\bfphi^{m+1}_j}_{j=1}^N$
\EndFunction
\end{algorithmic}
\end{algorithm}

\begin{algorithm}
    \begin{algorithmic}
        \caption{Check For Collisions}
        \Function{CheckForCollisions}{{$\set{\bfphi^{m}_j}_{j=1}^N$}, {$\set{\bfphi^{m+1}_j}_{j=1}^N$}}
        \State Calculate which $S_i^{m+1}$, $S_j^{m+1}$ are `close' to each other.
\For{\textbf{each} $S_i^{m+1}$, $S_j^{m+1}$ sufficiently close to each other}
\If{\texttt{CollisionDetected}($S^m_i$, $S^m_j$, $S^{m+1}_i$, $S^{m+1}_j$, 0, 1, $t_\text{hit}$)}
\State Interpolate the positions of  $\set{\bfphi^{m+1}_j}_{j=1}^N$ at time step $m + t_\text{hit}$
\State Split $\set{\bfphi^{m+1}_j}_{j=1}^N$ into a new pinned flow $\set{\hat \bfphi^{m+1}_j}_{j=1}^N$ 
\State Split $\set{\bfphi^{m+1}_j}_{j=1}^N$ into an unpinned flow $\set{\tilde \bfphi^{m+1}_j}_{j=1}^N$
\State $\set{\bfphi^{m+1}_j}_{j=1}^N \gets \set{\hat \bfphi^{m+1}_j}_{j=1}^N$
\State $\Phi \gets \Phi \cup \set{\set{\tilde \bfphi^{m+1}_j}_{j=1}^N } $
\EndIf
\EndFor

        \EndFunction
    \end{algorithmic}
\end{algorithm}

\begin{algorithm}
\begin{algorithmic}
\caption{Forced and Pinned MCF with Topological Change}
\label{MCF_Top_Change_Alg}
\Function{RunSimulation}{$\set{\bfphi^0_j}_{j=1}^N, t_\text{max} $}
\State $\Phi \gets (\set{\bfphi^0_j}_{j=1}^N)$ 
\For{$0 \leq m < t_\text{max}$}
\For{\textbf{each} flow $\set{\bfphi^m_j}_{j=1}^N$ in $\Phi$}
\State {$\set{\bfphi^{m+1}_j}_{j=1}^N$} =  \texttt{MoveNodesByForcedMCF}($\set{\bfphi^m_j}_{j=1}^N$)
\If{$\set{\bfphi^m_j}_{j=1}^N$ is pinned}
\State \texttt{CheckForCollisions}({{$\set{\bfphi^{m}_j}_{j=1}^N$}, {$\set{\bfphi^{m+1}_j}_{j=1}^N$}})
\EndIf
\State \texttt{Re-mesh}$\brac{\set{\bfphi^{m+1}_j}_{j=1}^N} $
\EndFor
\EndFor
\EndFunction
\end{algorithmic}
\end{algorithm}

\newpage
\bibliographystyle{alpha} 
\bibliography{dislocations.bib,curveshortening.bib,Plast.bib}

\end{document}